# Intrinsic Optical Bistability of Photon Avalanching Nanocrystals

Artiom Skripka[1,2], Zhuolei Zhang[1,3], Xiao Qi[1], Benedikt Ursprung[4], Peter Ercius[1], Bruce E. Cohen[1,5*], P. James Schuck[4*], Daniel Jaque[2*] and Emory M. Chan[1*]

[1]The Molecular Foundry, Lawrence Berkeley National Laboratory, Berkeley, California 94720, United States

[2]Nanomaterials for Bioimaging Group, Departamento de Física de Materiales, Facultad de Ciencias, Universidad Autónoma de Madrid, Madrid, 28049, Spain

[3]School of Chemistry and Chemical Engineering, Huazhong University of Science and Technology, 1037 Luoyu Road, Wuhan, 430074, China

[4]Department of Mechanical Engineering, Columbia University, New York, New York 10027, United States

[5]Division of Molecular Biophysics & Integrated Bioimaging, Lawrence Berkeley National Laboratory, Berkeley, California 94720, United States

*Corresponding authors: becohen@lbl.gov, p.j.schuck@columbia.edu, daniel.jaque@uam.es and emchan@lbl.gov

**Abstract**

Optically bistable materials respond to a single input with two possible optical outputs, contingent upon excitation history. Such materials would be ideal for optical switching and memory, yet limited understanding of intrinsic optical bistability (IOB) prevents development of nanoscale IOB materials suitable for devices. Here, we demonstrate IOB in $Nd^{3+}$-doped $KPb_2Cl_5$ avalanching nanoparticles (ANPs), which switch with high contrast between luminescent and non-luminescent states, with hysteresis characteristic of bistability. We elucidate a nonthermal mechanism in which IOB originates from suppressed nonradiative relaxation in $Nd^{3+}$ ions and from the positive feedback of photon avalanching, resulting in extreme, >200$^{th}$-order optical nonlinearities. Modulation of laser pulsing tunes hysteresis widths, and dual-laser excitation enables transistor-like optical switching. This control over nanoscale IOB establishes ANPs for photonic devices in which light is used to manipulate light.

## Main text

Optical bistability is a phenomenon in which a material can produce and switch between two different optical responses under a single input condition[1]. For example, an optically bistable material can exhibit two different levels of luminescence[2], reflectivity[3], transmission[4], or refractive index[5] when excited at the same power and wavelength. The two optical states of such materials are characterized by abrupt transitions between highly stable states, well-defined power thresholds, and hysteresis analogous to the flipping of magnetic domains[6]. Due to their ability to maintain discrete, photoswitchable, and stable states, optically bistable materials hold promise for optical logic, memory[7,8], and computing[9,10].

For practical application, bistable optical switches must be fast, reproducible, power-efficient, and readily integrated into dense circuitry. Most reports of optically bistable materials are based on bulk materials or utilize external cavities[3,4,11,12], entailing complex architectures and complicating potential device fabrication. Reducing the size of bistable materials below 100 nm would facilitate very-large-scale integration and solution-based processing. Such materials would ideally derive their optical bistability from purely electronic processes intrinsic to the host material, rather than poorly controllable thermal effects, but a limited understanding of the mechanisms for achieving non-thermal intrinsic optical bistability (IOB) has slowed the development of nanoscale IOB.

To discover nanoscale IOB materials, we hypothesized that recently described giant nonlinear optical responses of photon avalanching nanoparticles (ANPs)[13,14] could provide the positive feedback that is a critical requirement for observing IOB[15,16]. While IOB has been previously observed in $Ln^{3+}$-doped bulk crystals[17–20] and nanoparticles[21,22], the mechanisms in these materials were ultimately attributed to uncontrollable thermal phenomena[2,23]. Here, we demonstrate nonthermal IOB in $Nd^{3+}$-doped $KPb_2Cl_5$ nanoparticles and elucidate a mechanism in which IOB is driven by a photon avalanche (PA) with an extreme nonlinearity order $s > 200$, where the emission intensity $I_{em}$ scales with the excitation power $P$ as $I_{em} \propto P^s$. These low-phonon-energy ANPs exhibit two luminescent states – a dark, non-emissive state and a bright state that emits upconverted light (Fig. 1A) – with the current ANP state determined by its 1064-nm excitation intensity history. We find that the dark state is stabilized by nonradiative quenching of excited energy levels, and the bright state is stabilized by positive feedback from a PA process (Fig. 1B). By moving between these states and manipulating the hysteresis that controls their bistability, we can efficiently switch ANP luminescence and demonstrate high-contrast, transistor-like optical behavior.

## Results and Discussion

### Bistable luminescence of ANPs

To identify nanomaterials that exhibit extreme optical nonlinearities, and possibly IOB, we explored $KPb_2Cl_5$ nanocrystals because their low phonon energies enable PA in $Nd^{3+}$ dopants at room temperature (see Supplementary Information for synthesis and characterization). Additionally, the $KPb_2Cl_5$ matrix provides chemical stability at both ambient and cryogenic conditions[24]. We sought to determine if the nonlinearities of these materials would be more pronounced when cooled since multiphonon relaxation (MPR) rates theoretically decrease by several orders of magnitude at low temperature (Fig. S3)[25]. Liquid nitrogen-cooled films of $KPb_2Cl_5$ nanoparticles doped with 16 mol% $Nd^{3+}$ show sharp increases in 810 nm emission when the 1064 nm pump intensity reaches a threshold value of 6.7 $kW \cdot cm^{-2}$ (Fig. 2a). By fitting the linear region of the power-dependent luminescence curve at the switch-on threshold, we measure a slope indicating a nonlinearity of $s \geq 200$ (Fig. 2a), constrained by instrumental limitations and significantly steeper than nonlinear reports of PA at room temperature. Even when stepping pump intensities at the smallest possible increments, we do not detect emission between discrete dark and bright states. Such discrete jumps may reflect the threshold behavior of individual ANPs, since the spatial resolution at such extreme nonlinearities is comparable to the size of individual ANPs[13,26]. These results show that, at 77 K, $Nd^{3+}$-doped $KPb_2Cl_5$ nanocrystals exhibit extreme photon avalanching that is a prerequisite for IOB.

Reduction of the pump intensity below the 6.7 $kW \cdot cm^{-2}$ switch-on threshold does not reduce ANP emission significantly until the irradiance decreases to 4.2 $kW \cdot cm^{-2}$, at which point the ANPs abruptly turn dark (Fig. 2a). This non-zero difference between excitation intensity required to switch the ANPs on and off creates a hysteresis in their optical response, indicating optical bistability. Within this bistable region, the two stable states of the ANPs are manifested as dark or bright emission at intermediate pump intensity (e.g., 5 $kW \cdot cm^{-2}$), depending on whether the luminescence is measured before or after reaching the switch-on threshold. Such bistability is spectrally observed in multiple $Nd^{3+}$ emission lines, including the 810 nm ($^4F_{5/2} \rightarrow {}^4I_{9/2}$) and 880 nm ($^4F_{3/2} \rightarrow {}^4I_{9/2}$) transitions (Fig. 2b).

We observed this luminescence hysteresis on two different optical setups (Fig. S4) and on different substrates (Fig. S5). Furthermore, when we varied the acquisition dwell times (0.1 – 10 s) used for each pump power, the hysteresis widths and thresholds remained constant (Fig. S6), indicating that the hysteresis does not originate from time-dependent factors such as laser-induced heating or slow PA rise-times (ca. 20 ms, Fig. S21). Even room temperature rise times (50-70 ms)[24] are shorter than the shortest dwell times (0.1 s), ensuring that samples are measured at steady state. All of these observations support our conclusion that the ANP emission signal is intrinsically bistable and not the result of experimental artifacts[27].

To obtain statistical information about the reproducibility of IOB and characteristics including switch-on threshold, hysteresis width, and intensity contrast, we measured the pump power dependence at multiple spots across a film of $KPb_2Cl_5:Nd^{3+}$ ANPs (Fig. S7) (Fig. 2c). The average switch-on power density is 5.2 $kW \cdot cm^{-2}$, corresponding to 64 μW, suitable for low-power requirements (μW – mW) in optical computing[28]. On average, the width of luminescence hysteresis is 1 $kW \cdot cm^{-2}$, and the photoswitching contrast of the $KPb_2Cl_5:Nd^{3+}$ ANPs (i.e., signal-to-noise ratio at the switch-on threshold) is 14±4 dB, sufficiently high to unambiguously differentiate between bright and dark ANP states[29]. This contrast is one to two orders of magnitude greater than that of previously reported IOB nanomaterials, which were studied at RT (see Supplementary Table S2 for a comparison of $KPb_2Cl_5:Nd^{3+}$ ANPs to other optically bistable materials).

To demonstrate the stability and reproducibility of this IOB, we repeated the optical hysteresis measurements on the same spot (Fig. 2a) 100 times over 12 hours (Fig. S8). Despite minor variations in the switch-on threshold (6.52 ± 0.21 $kW \cdot cm^{-2}$) and hysteresis width (2.20 ± 0.15 $kW \cdot cm^{-2}$), possibly due to sample drift, the IOB is consistently observed for every cycle. Notably, these ANPs show no photodarkening or blinking[30,31].

To probe how these nanocrystals give rise to IOB, we varied their $Nd^{3+}$ doping levels. The high nonlinearity and luminescence hysteresis are predominantly observed in samples with $Nd^{3+}$ concentrations above 4 mol% (Fig. S9), suggesting that the nonlinearity requisite for the IOB is driven by a mechanism such as PA that incorporates distance-dependent energy transfer processes[13,32]. To understand the role of the host matrix, we measured the luminescence power dependence of $Nd^{3+}$ ions doped in a higher phonon energy host, i.e., $NaYF_4$. However, we do not observe PA or luminescence hysteresis even when $NaYF_4$: $Nd^{3+}$ nanocrystals are cooled to 8 K (Fig. S10), underscoring the need for low-phonon-energy hosts to suppress MPR and observe PA and IOB. Thus, cryogenic temperatures alone are insufficient to promote IOB from nonlinear $Nd^{3+}$ luminescence.

We further investigated the impact of thermal phonons on the quenching of $Nd^{3+}$ energy levels (e.g., $^4I_{11/2}$) by measuring the power-dependent luminescence of $KPb_2Cl_5:Nd^{3+}$ ANPs at different temperatures. We find that higher temperatures lead to a lower degree of nonlinearity and narrowing of luminescence hysteresis (Fig. 2d), making thermal contributions to IOB unlikely.

Importantly, we observe IOB at temperatures as high as 150 K (Fig. 2d). The switch-on threshold also shifts to lower power densities with increasing temperatures, correlating with an enhanced phonon-assisted ground state absorption (GSA). Notably, at 160 K, ANPs do not show luminescence hysteresis despite their extremely nonlinear ($s$ = 70) luminescence, suggesting that extreme nonlinearities ($s \geq 100$) are required to induce IOB in $KPb_2Cl_5:Nd^{3+}$ ANPs.

**Mechanism of IOB in ANPs**

To corroborate these experimental results and better understand this observed IOB, we used a differential rate equation model[33] to numerically simulate the time-dependent population of the $4f^N$ energy levels of $Nd^{3+}$ ions in a $KPb_2Cl_5$ host (see Supplementary Note S1 for details). Our simulations reproduce the nonlinear power scaling of $KPb_2Cl_5:Nd^{3+}$ luminescence and, most importantly, the luminescence hysteresis of experimental observations (Fig. S11). Similar results were also found using a simplified rate equation model (see Supplementary Note S2). These simulations do not incorporate thermal phenomena, highlighting that IOB in $KPb_2Cl_5:Nd^{3+}$ ANPs is governed only by the optical activation of electronic 4f energy levels, which distinguishes this system from thermal IOB in other materials[19,23].

The photophysical mechanism (Fig. S13) extracted from the simulations[33] resembles the canonical PA process in which excited state absorption (ESA, thick orange arrow in Fig. 3a) dominates over ground state absorption (GSA, thin orange arrow) and couples to a cascade of CR processes (dashed arrows, CR1-5) to form an "energy loop." Each CR process multiplies the population of low-lying excited manifolds of $Nd^{3+}$ ions, like $^4I_{11/2}$, increasing the rate of ESA and CR from those levels. Repeated CR+ESA amplifies those populations exponentially in a positive feedback loop, resulting in the observed nonlinear behavior.

Such looping is supported by our observation that simulated populations of the $Nd^{3+}$ ground ($^4I_{9/2}$) and first excited ($^4I_{11/2}$) 4f electronic levels are inverted at the switch-on pump threshold (Fig. 3b). Notably, this population inversion (PI) is maintained even when the power is reduced below that threshold, resulting in hysteresis in the power dependence of the population ratio between the $^4I_{11/2}$ and $^4I_{9/2}$ energy levels. The PI is correlated with the emergence of the bright state of ANPs, suggesting that bright and dark steady states of these bistable ANPs are distinguished by their inverted and non-inverted excited level populations, respectively.

To identify the critical transitions responsible for IOB, we performed a series of "knock out" simulations in which we systematically deactivate individual cross-relaxation pathways in the CR cascade (CR1-5) that drives the PA (see SI for details)[34]. When MPR and other parameters are held constant, the simulated power-dependent luminescence of knock-out systems reveals that among the cross-relaxation processes, the CR1 in Fig. 3a [($^4I_{9/2}$→$^4I_{11/2}$):($^4I_{13/2}$→$^4I_{11/2}$)] has the most impact on the performance of the bistable ANPs. Deactivating CR1 reduces the nonlinearity by 5.8-fold (to *s* = 17) and eliminates luminescence hysteresis (Fig. 3c). Although CR processes from the $^4F_{3/2}$ excited energy level (CR3-5 in Fig. 3a) are not essential to establish PI, knocking them out shifts the switch-on threshold to higher values (Fig. S11). We believe these pathways are needed to repopulate lower $Nd^{3+}$ energy levels ($^4I_{11/2}$, $^4I_{13/2}$, $^4I_{15/2}$) after photon absorption and contribute to the nonlinear behavior of ANPs through a CR cascade. Finally, we compared the rate of CR1 in the wild-type (WT – no knock-outs) $KPb_2Cl_5:Nd^{3+}$ to that of other transitions. We find that within the bistable region of luminescence, the $^4I_{11/2}$ energy level is populated via CR1 faster than it is depopulated via MPR (Fig. S14 and S15); this dominance of CR1 over quenching processes appears to be a necessary condition for IOB, which is satisfied in low-phonon-energy matrices that result in low MPR rates (Supplementary Table S4).

Based on these mechanistic observations, we construct a conceptual scheme explaining how IOB occurs in $KPb_2Cl_5:Nd^{3+}$ ANPs (Fig. 1b): these bistable ANPs exhibit two stable macroscopic states: (1) a dark, non-emissive state with $Nd^{3+}$ ions occupying their ground energy level and (2) an avalanching, luminescent state with $Nd^{3+}$ ions populated predominantly in their first excited energy level. At low 1064 nm pump intensities ($P_{low}$ in Fig. 1b), the dark state ① is stabilized from moderate increases in pump power by the suppression of non-resonant GSA in the low-phonon-energy $KPb_2Cl_5$ host at low temperatures. MPR also stabilizes the dark state by quenching the few excited $Nd^{3+}$ ions before they can achieve the intermediate $^4I_{11/2}$ populations necessary for avalanching. The bright state (③) – established at high intensities ($P_{high}$) – is stable because the positive feedback of the avalanching process maintains PI. Even when the pump power is reduced below the PA threshold, avalanching persists because the intermediate $^4I_{11/2}$ population has already been seeded and is not readily quenched due to suppressed MPR in the low-phonon-energy host. Due to the giant nonlinearity of PA, only a small amount of ESA from this $^4I_{11/2}$ level is necessary to sustain the cascade of CR back to the same level and counteract its depopulation by MPR. As a result, both the bright and dark states are stable at moderate pump intensities ($P_{med}$, ②).

To reiterate, this IOB is a direct consequence of the low phonon energy of $KPb_2Cl_5$ and the positive feedback of PA in $Nd^{3+}$ dopants. These driving forces create a potential energy barrier (i.e., instability region) that separates the two states in the bistable regime (see potential energy surface in Fig. 1b). This barrier ultimately gives rise to the latching hysteresis and history dependence that characterizes the bistable luminescence of $KPb_2Cl_5:Nd^{3+}$ ANPs.

**Temporal modulation of IOB**

After establishing that IOB in $KPb_2Cl_5:Nd^{3+}$ ANPs is realized through a PA mechanism, we hypothesized that the dynamics of this positive feedback could be manipulated by modulating the excitation power over time, potentially influencing the IOB behavior. To test this hypothesis, we performed power dependence studies with a 1064 nm pump modulated at various frequencies and duty cycles. We find that modulating the laser selectively inhibits or induces IOB in ANPs and allows for control over luminescence hysteresis (Fig. 4, Fig. S16 and S17). Slow pulsing (<170 Hz, 40% duty cycle) of the excitation laser eliminates the hysteresis observed with continuous wave (CW) 1064 nm excitation and decreases the PA nonlinearity significantly ($s$ = 40 at 140 Hz, Fig. 4a). In contrast, at higher pulsing frequencies (and lower duty cycles, see SI), the hysteresis widens (Fig. 4b), and the PA threshold shifts to higher average powers (Fig. 4c). At 1000 Hz pump frequency, the hysteresis width reached 4.5 $kW{\cdot}cm^{-2}$ (Fig. 4) and 24 $kW{\cdot}cm^{-2}$ (Fig. S18) for 40% and 10% pulse duty cycle, respectively, significantly wider than under CW excitation (1.6 $kW{\cdot}cm^{-2}$).

We rationalize this dynamic control over IOB by considering the pulse duration and period between pulses. The greater switch-on power thresholds at higher frequencies arise due to the decreased probability of GSA with decreasing pulse energies (25 nJ at 100 Hz vs. 3.6 nJ at 1000 Hz). Similarly, IOB emerges at high pulsing frequencies because the time period between pulses is short enough to sustain the positive feedback of CR+ESA in $KPb_2Cl_5:Nd^{3+}$ ANPs. We also examined the notable difference in hysteresis characteristics between CW and high-frequency pulsed excitation of ANPs. The widening of hysteresis and higher average switch-on powers observed in the 500-1000 Hz range are not observed at pulse frequencies higher than 4000 Hz because such closely spaced pulses become analogous to quasi-CW excitation (Fig. S19 and S20). Based on the frequencies at which ANPs begin to exhibit luminescence hysteresis (180 Hz and 300 Hz at 40% and 10% duty cycle, respectively), we estimate that the PI in ANPs lasts up to 3 ms, which is likely related to the lifetime of the $^4I_{11/2}$ excited energy level of $Nd^{3+}$ ions in $KPb_2Cl_5$ matrix (2.3 ms at room temperature[35]). The duration of PI also dictates the photoactivation kinetics of the $KPb_2Cl_5:Nd^{3+}$ ANPs. With high-frequency pulsing, we measure short rise times for emitting energy levels populated directly by ESA (e.g., $^4F_{3/2}$ is populated in 305 μs). In contrast, PI is lost between low-frequency excitation pulses, resulting in millisecond rise times from the initially slow GSA followed by PA, similar to $KPb_2Cl_5:Nd^{3+}$ ANPs at RT[24]. See Supplementary Note S3 for details.

The fact that this IOB occurs via photophysical interactions between dopant ions (i.e., nonthermally) allows us to actively control the switch-on power threshold, hysteresis width, and nonlinearity of bistable ANPs, underscoring their utility in photonic applications[36–38].

**Optical switching via instability crossing**

With the ability to induce and control IOB in $KPb_2Cl_5:Nd^{3+}$ ANPs, we sought to demonstrate how their bistable response can be used for switching logic and memory. In principle, a bistable ANP can be treated as a bit whose dark state (i.e., a value of 0) may be flipped by increasing the pump intensity to activate the ANP's bright (1) state. The bright bit can then be maintained at lower pump intensities. However, this single-laser approach may be impractical as it requires varying the pump intensity and a dedicated laser beam to address each bit independently.

To overcome these challenges, we hypothesized that bistable ANPs could be manipulated using laser pumps at two different wavelengths, mimicking a transistor (Fig. 5a). We developed an approach that uses an 808 nm control laser pulse to set an ANP in its bright state and a constant 1064 nm laser as a bias to maintain this state. The advantage of this two-laser approach is that the 1064 nm bias laser, in principle, can be spread diffusely to maintain the state of many bits simultaneously, allowing the control laser to raster freely and flip individual bits. Furthermore, the 808 nm control pulse enables fast switching of ANPs from their dark to a bright state because it is resonant with GSA ($^4I_{9/2} \rightarrow {}^4F_{3/2}$) in $Nd^{3+}$ ions. This resonant excitation of the ANP bright states at 808 nm bypasses, or crosses, the instability barrier that normally exists when exciting ANPs non-resonantly at 1064 nm (Fig. 5b). Such "instability crossing" simplifies bit operations since it eliminates the need to vary the 1064 nm laser power.

When we implemented this two-laser scheme on a film of ANPs, we observed that their 880 nm ($^4F_{3/2} \rightarrow {}^4I_{9/2}$) luminescence flips from a dark (0) to a bright (1) state after a ~1 s pulse of the 808 nm laser and is maintained as long as a 1064 nm bias laser (4.7 $kW \cdot cm^{-2}$) is present (Fig. 5c). Notably, the switching between the two stable states is achieved at extremely low 808 nm power density (7 $W \cdot cm^{-2}$, corresponding to 75 nW power). We note that the latching of the bright state via instability crossing would not occur with thermally activated avalanching and is enabled by the PA-driven IOB in $KPb_2Cl_5:Nd^{3+}$ ANPs.

To determine whether photoswitching occurs only when ANPs are biased within the bistable region, we performed control experiments by applying the 1064 nm laser bias outside the bistable region and observed no change in ANP luminescence before or after the 808 nm input (Fig. 5c). Furthermore, we demonstrate the instability crossing using a pulsed 1064 nm laser to exploit larger hysteresis (Fig. S22) and using control lasers with wavelengths resonant with different GSA transitions of $Nd^{3+}$ ions (875 nm for $^4I_{9/2} \rightarrow {}^4F_{3/2}$ GSA and 745 nm for $^4I_{9/2} \rightarrow {}^4F_{7/2}$ GSA, see Fig. S23). These findings demonstrate the versatility of bistable ANPs and establish the foundation for their use as nanoscale optical transistors and volatile optical memory[39].

**Conclusion**

Our findings demonstrate the realization of purely optical, nonlinear, and intrinsically bistable luminescence in $KPb_2Cl_5:Nd^{3+}$ nanocrystals. This behavior is realized by suppressing phonon-mediated transitions and through the positive feedback of photon avalanche among $Nd^{3+}$ dopants. Unlike other $Ln^{3+}$-doped materials that exhibit intrinsic optical bistability, IOB in $KPb_2Cl_5:Nd^{3+}$ ANPs arises from a nonthermal mechanism quantitatively explained by photophysical modeling. The lack of thermal contribution to IOB in our model is supported by multiple lines of experimental evidence, including (1) the continued observation of IOB at low duty cycles, the insensitivity of the hysteresis to (2) variations in scan rate or (3) the thermal conductivity of the substrate, and (4) the ability to photoswitch bistable particles with a low-power laser. These observations – discussed in greater detail in Supplementary Discussion S1 – are fully consistent with an all-optical mechanism for IOB and are difficult to reconcile with thermally mediated bistability.

The all-optical nature of IOB in $KPb_2Cl_5:Nd^{3+}$ ANPs allows tuning of luminescence hysteresis characteristics, optical switching, and memory. IOB-based switching of $KPb_2Cl_5:Nd^{3+}$ ANPs is complementary to, but distinct from, the previously reported photoswitching of $NaYF_4:Tm^{3+}$ ANPs. While photodarkening of $Tm^{3+}$-doped ANPs occurs through the formation of long-lived defects akin to persistent memory, IOB-based photoswitching is a dynamic process analogous to volatile memory[39], as we demonstrate here. We envision that bistable ANPs can address several challenges that currently limit the construction of digital optical logic gates, such as cascadability, logic-level restoration, and the absence of critical biasing[10,40]. Combined with solution processing and direct lithography methods, 3D volumetric interconnects of bistable ANPs could be fabricated for high-density optical computing[10,41]. Furthermore, bistable ANPs can be operated with low-power, cost-efficient lasers compatible with telecommunication wavelengths or silicon detectors, facilitating integration with existing technologies. These bistable ANPs are analogous to nanoscale optical transistors, paving the way for general-purpose digital optical computing, neuromorphic circuitry, imaging, and related photonic technologies.

## Acknowledgments

A.S. acknowledges support from the European Union's Horizon 2020 research and innovation program under the Marie Skłodowska-Curie Grant Agreement No. 895809 (MONOCLE). X.Q., B.E.C., and E.M.C. were supported by the Defense Advanced Research Projects Agency (DARPA) under Contract No. HR0011257070. Z.Z. and E.M.C. were supported by DARPA Contract No. HR001118C0036. C.L and P.J.S. acknowledge support from DARPA Contract No. HR00112220006 and from the National Science Foundation grant CHE-2203510. Work at the Molecular Foundry was supported by the U.S. Department of Energy (DOE) under Contract No. DE-AC02-05CH11231 through the Office of Science, Office of Basic Energy Sciences (BES). Electron microscopy (X.Q., P.E., E.M.C.) was funded in part under the same DOE contract through the BES Materials Sciences and Engineering Division within the KC22ZH program. Work at Universidad Autónoma de Madrid (D.J.) was financed by the Spanish Ministerio de Innovación y Ciencias under Project No. NANONERV PID2019-106211RB-I00 and by the Comunidad Autónoma de Madrid (S2022/BMD-7403 RENIM-CM).

## Author Contributions Statement

A.S., D.J., and E. M. C. designed and conceptualized the research. B.C., P.J.S., D.J., and E. M. C. directed the study. A.S. performed optical characterization experiments, numerical simulations, and data analysis. X. Q. and Z. Z. synthesized the nanoparticles. X.Q. and P.E. performed electron microscopy imaging and analysis. B. U. automated optical data acquisition. A.S. and E. M. C. wrote the original manuscript with B.C., P. J. S., and D. J. All authors discussed the results and revised the manuscript.

## Competing Interests Statement

The authors declare the following competing interests: A. S., Z. Z., B. C., P. J. S, and E. M. C. have submitted a United States patent application under review that describes the synthesis and application of low phonon energy nanoparticles based on alkali lead halides.

## Figures

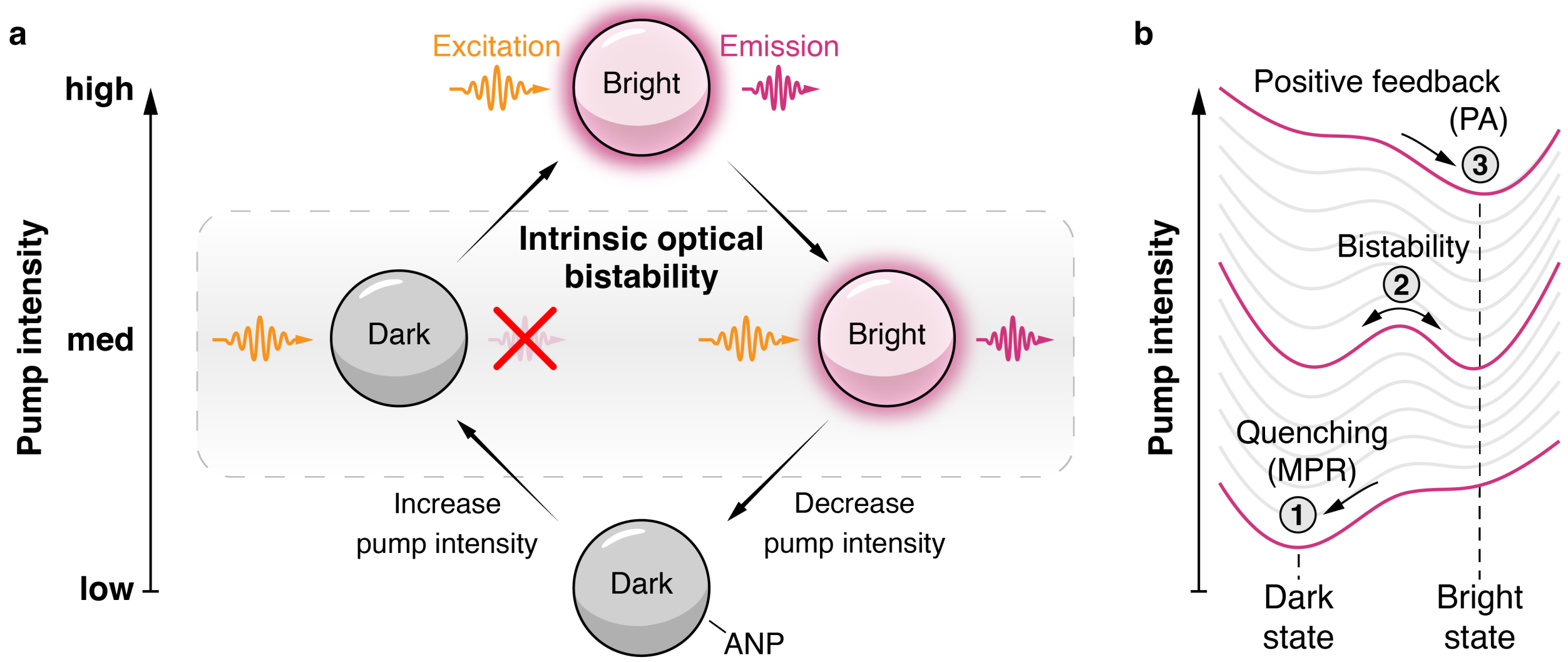

**Fig. 1**. **Intrinsic optical bistability in $KPb_2Cl_5:Nd^{3+}$ ANPs**. **a**, Bistable ANPs switch from dark to bright, stable state at high 1064 nm pump intensity, which persists even after the power is decreased. ANPs that have not been previously subjected to high intensities show much weaker emission under the same excitation conditions. **b**, Landscape of ANP states at different pump intensities. Quenching by multiphonon relaxation (MPR) stabilizes the dark state (1). In contrast, the positive feedback loop of excited state absorption (ESA) and cross-relaxation (CR) leads to photon avalanching (PA) and stabilizes the bright state (3). Intermediate excitation intensities result in a double well in which the bistable dark and bright states are separated by a potential energy barrier that gives rise to instability and history dependence.

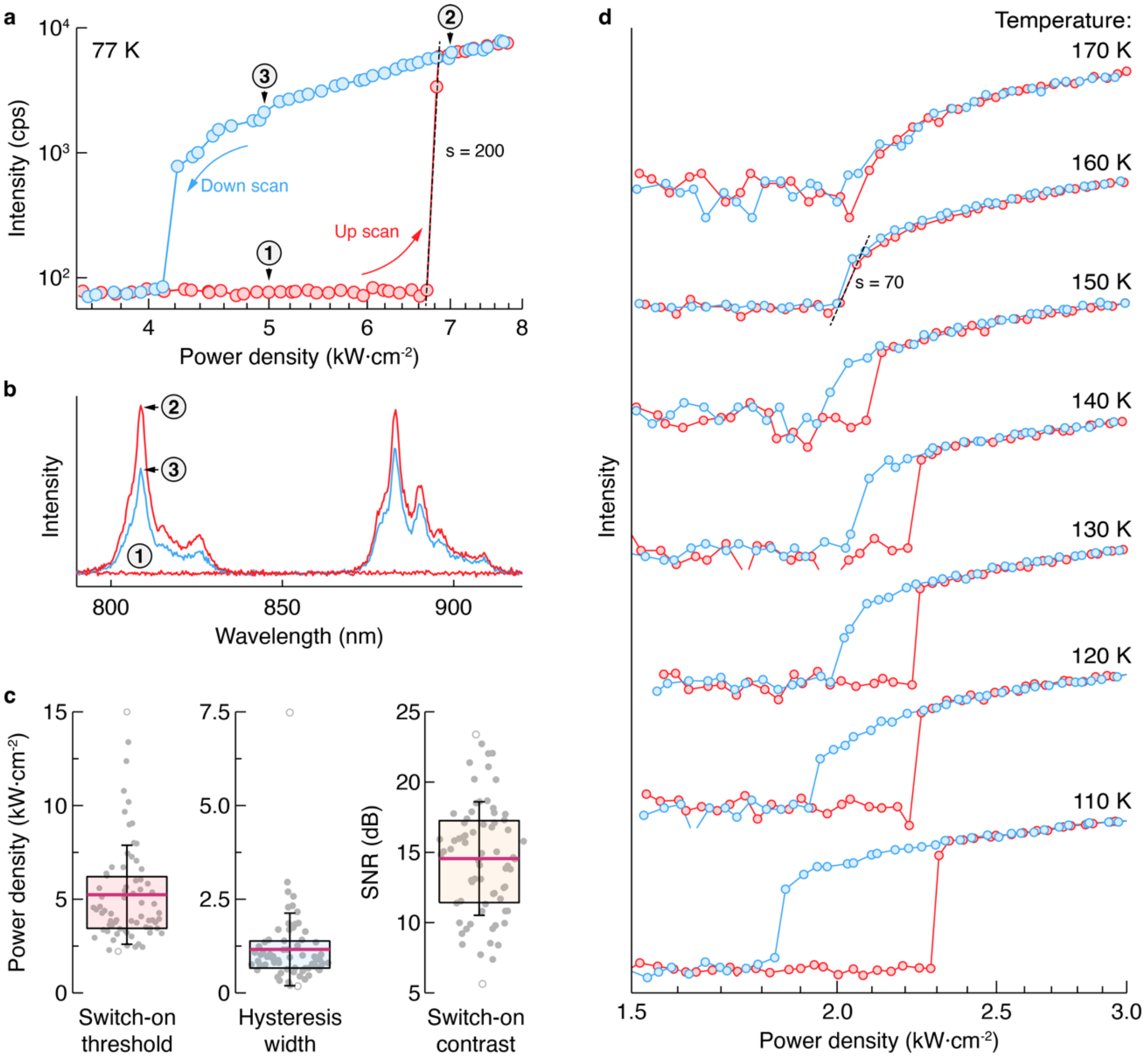

**Fig. 2**. **Bistable luminescence of $KPb_2Cl_5:Nd^{3+}$ ANPs.** **a**, Pump power dependence of $Nd^{3+}$ emission (810 nm, $^4F_{5/2} \rightarrow {}^4I_{9/2}$) in $KPb_2Cl_5:Nd^{3+}$ ANPs at 77 K. Red data points represent luminescence intensity acquired with increasing excitation intensity (up scan), and blue with decreasing (down scan). Number indices represent the sequence of events. **b**, Luminescence spectra of $KPb_2Cl_5:Nd^{3+}$ nanocrystals corresponding to power density and scan direction labeled in **a** by the number indices. **c**, Switch-on threshold, hysteresis width, and switch-on contrast from *n* = 71 independent measurement spots on a film sample of $KPb_2Cl_5:Nd^{3+}$ ANPs. Dots – experimental data points. Box plots display the mean (pink line), standard deviation (whiskers), and 25$^{th}$ and 75$^{th}$ percentiles (box bounds). Minima and maxima are shown as open circles. Measurement parameters are indicated below each box plot. **d**, Pump power dependence of $Nd^{3+}$ emission (810 nm, $^4F_{5/2} \rightarrow {}^4I_{9/2}$) in $KPb_2Cl_5:Nd^{3+}$ ANPs in the 110-170 K temperature range. Emission scaling at the switch-on threshold is derived from a linear fit of the log(light-light) curves; see dashed lines and nonlinearity, s, values in **a** and **d**.

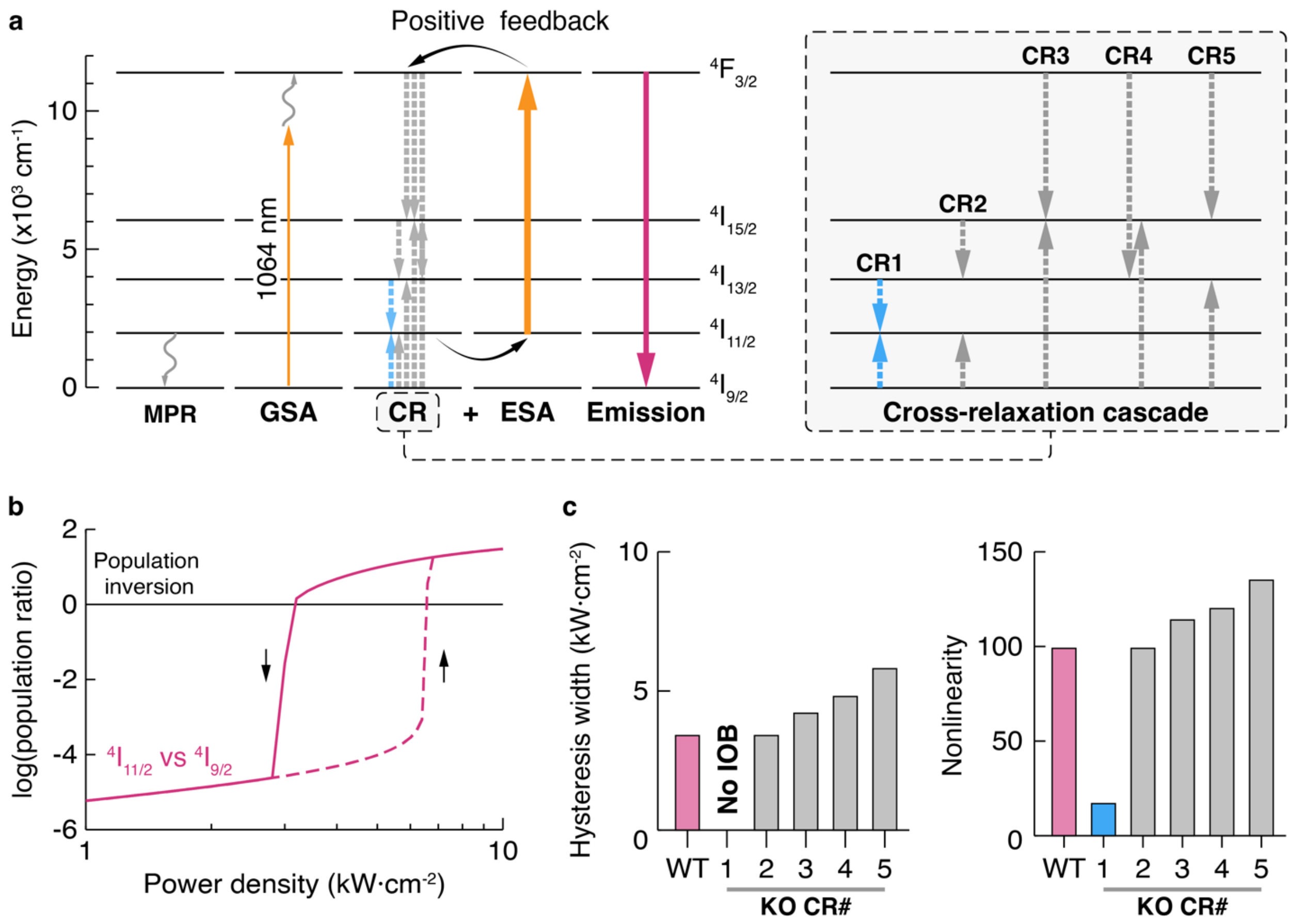

**Fig. 3**. **Model of IOB in $KPb_2Cl_5$:$Nd^{3+}$ ANPs. a**, Simplified energy level diagram of $Nd^{3+}$ ions showing GSA, cross-relaxations (CR#), ESA, and emission. ESA and cross-relaxations form a positive feedback loop, leading to giant signal amplification and IOB. Competing MPR (wavy arrow) is also shown. **b**, Pump power dependence of $Nd^{3+}$ excited ($^4I_{11/2}$) vs. ground ($^4I_{9/2}$) energy level population ratio in a simulated $KPb_2Cl_5$: 20 mol% $Nd^{3+}$ system. The dashed line corresponds to power up and solid to power down scans. **c**, Influence of cross-relaxation pathway knock-outs (KO) on the hysteresis width and nonlinearity for $^4F_{3/2} \rightarrow {}^4I_{9/2}$ radiative transition. WT – wild type, CR# – as shown in **a**.

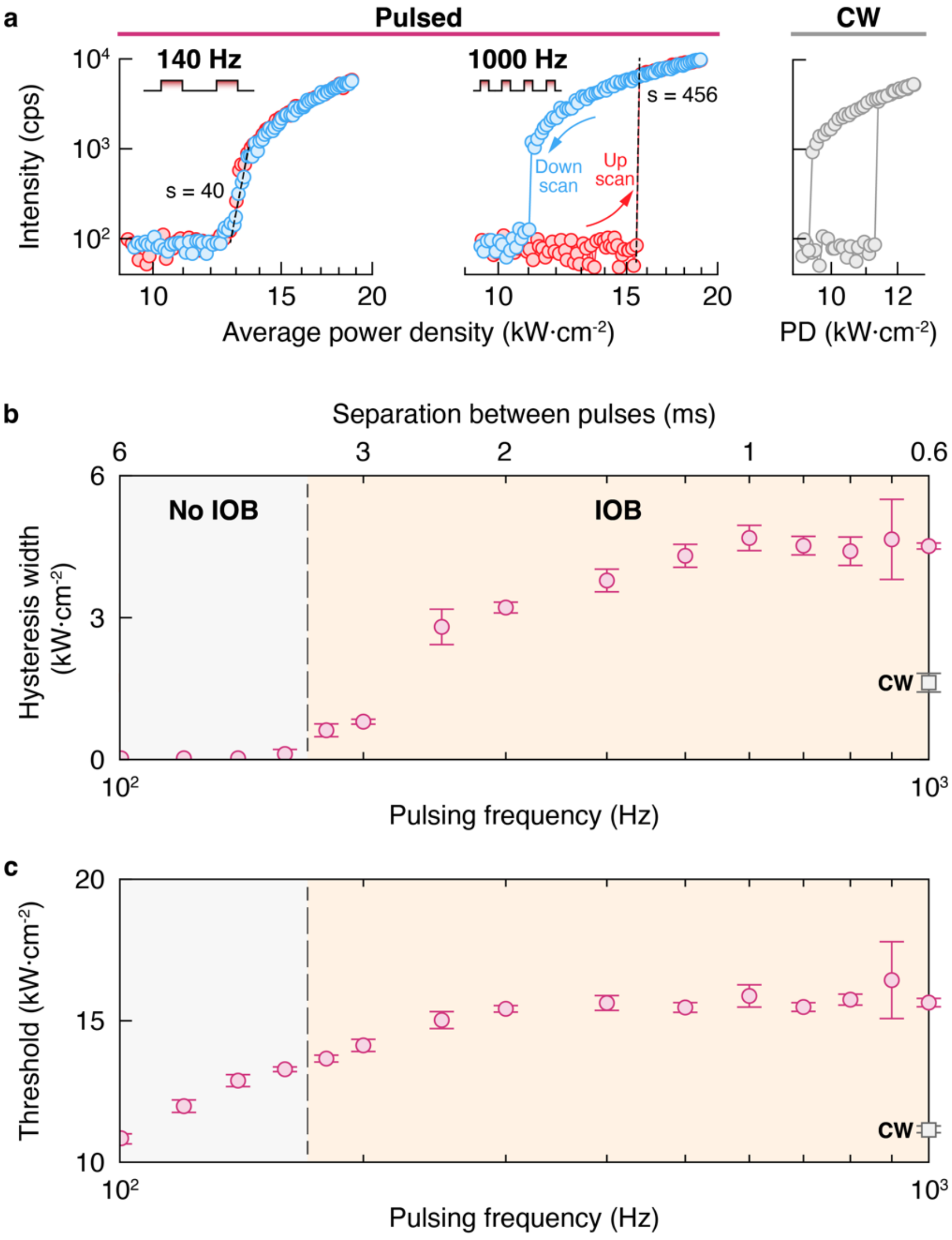

**Fig. 4. Temporal modulation of luminescence hysteresis. a**, Pump power dependence of $Nd^{3+}$ emission in $KPb_2Cl_5:Nd^{3+}$ ANPs at 77 K under 1064 nm continuous wave (CW) and pulsed pump excitation (40% duty cycle). **b**, hysteresis width and **c**, PA switch-on threshold vs. pump pulse frequency (40% duty cycle). Corresponding dependence on the temporal separation between pulses is also shown. Respective values under the CW pump are shown as squares. Emission scaling in **a** is derived from a linear fit of the log(light-light) curves. Data in **b** and **c** are shown as the mean values ± 1 standard deviation ($n$ = 3).

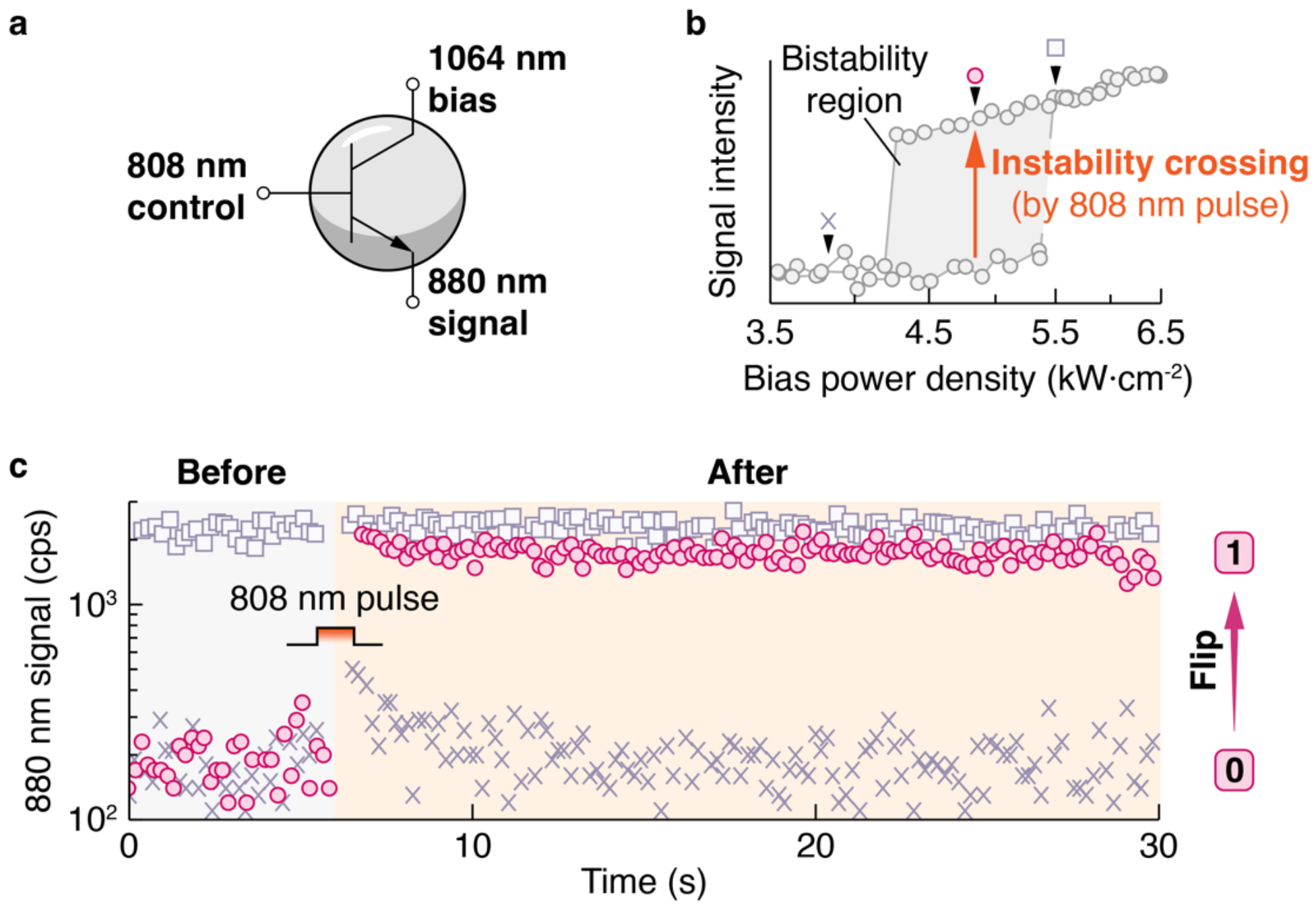

**Fig. 5**. **Photoswitching of ANPs via instability crossing. a**, Schematic representation of transistor-like addressing of $KPb_2Cl_5:Nd^{3+}$ nanocrystals with 808 nm control input (7 W·cm$^{-2}$) under a constant 1064 nm pump bias (4.7 kW·cm$^{-2}$). **b**, Pump power dependence of $KPb_2Cl_5:Nd^{3+}$ nanocrystals used for photoswitching experiments. The bistability region (gray) and instability crossing (orange arrow) are shown. The 1064 nm bias for the experiment in **c** was applied at three different power densities: a dark state outside the bistability region (cross), within the bistability region (circle), and a bright state outside the bistability region (square). **c**, Time recording of emission at 880 nm of $KPb_2Cl_5:Nd^{3+}$ ANPs before and after 808 nm control pulse when biased at 1064 nm pump. The symbols in **c** indicate the 1064 nm bias power densities for each trace as defined by the placement of those symbols in **b**.

## Methods

All experimental details, including the synthesis of nanocrystals, structural characterization, optical measurements, and simulations, are provided in the Supplementary Information.

## Data availability

All data in the main text or the supplementary materials are available from the corresponding author upon reasonable request.

## Code availability

All code used in this paper are available from the corresponding author upon reasonable request.

## Supplementary Information

## Table of Contents

# 1. Materials and Methods

## 1.1 Materials

Lead acetate trihydrate ($Pb(CH_3COO)_2{\cdot}3H_2O$, 99.99%) was purchased from Alfa Aesar, Potassium carbonate, anhydrous ($K_2CO_3$, 99%), Neodymium (III) acetate hydrate ($Nd(CH_3COO)_3{\cdot}xH_2O$, 99.9%), Neodymium (III) chloride ($NdCl_3$, 99.9+%), Yttrium (III) chloride ($YCl_3$, 99.9+%), myristoyl chloride (97%), ammonium fluoride ($NH_4F$, 99.9%), oleic acid (OA, 90%), octadecene (ODE, technical grade, 90%), and hexane (anhydrous, 99.5%) were purchased from Sigma-Aldrich. Sodium oleate (>97%) was purchased from TCI. All chemicals were used without any further purification.

## 1.2 Synthesis of $KPb_2Cl_5:Nd^{3+}$ nanocrystals

$KPb_2Cl_5:Nd^{3+}$ nanocrystals were synthesized according to the procedure by Zhang et al.[1] In a typical synthesis, $K_2CO_3$ (6.9 mg), $Pb(CH_3COO)_2{\cdot}3H_2O$ (76 mg), 3 mL of OA, 0.5 mL of OM, $Nd(CH_3COO)_3{\cdot}xH_2O$ (0.25 mmol), and 6 mL of ODE were added into a 50 mL 3-neck round bottom flask. The solution was stirred and degassed on a Schlenk line at 40 °C for 5 min and then heated to 120 °C degassing for 1 hour. Subsequently, the temperature was increased to 260 °C under $N_2$. Upon reaching this temperature, 0.3 mL of myristoyl chloride precursor was swiftly injected. The reaction flask was immediately immersed in an ice−water bath. Finally, after adding 5 mL of hexane, the crude solution was centrifuged at 3000 rpm for 5 min. The supernatant was discarded, and the precipitate was redispersed in 5 mL of hexane for further use. $Nd^{3+}$-doped $KPb_2Cl_5$ NPs with varying $Nd^{3+}$ concentrations were synthesized following the method described above, except the nominal $Nd^{3+}$ amount was varied using values of 0.004, 0.01, 0.04 and 0.10 mmol $Nd(CH_3COO)_3{\cdot}xH_2O$.

## 1.3 Synthesis of $NaYF_4:Nd^{3+}$ nanocrystals

$NaYF_4:Nd^{3+}$ nanoparticles were synthesized using a previously described synthesis[2], with some modification. To a dry 100 mL 3-neck round bottom flask (0.9 mmol, 175.7 mg) of $YCl_3$ and (0.1 mmol, 25.1 mg) of $NdCl_3$ were added together with OA (6 mL) and ODE (14 mL). The flask was stirred, placed under vacuum, and heated to 100 ºC for 1 hour, causing the solution to become clear. The flask was then filled with $N_2$, and sodium oleate (2.5 mmol, 761.1 mg) and $NH_4F$ (4 mmol, 148.1 mg) were added. The flask was subsequently placed under vacuum and stirred for another 20 min, followed by $N_2$ flushing 3 times. The reaction was heated to 320 ºC and allowed to react under $N_2$. After 40 min of reaction time, the flask was rapidly cooled down to room temperature by a strong stream of air, and nanoparticles were isolated with the help of EtOH (20 mL) and centrifugation (3000 x g, 5 min). The nanoparticles were washed twice with hexane:EtOH (1:1 *v*/*v*) and redispersed in 4 mL of hexane for storage.

## 1.4 Structural characterization

Transmission electron microscopy (TEM) and high-angle annular dark-field scanning TEM (HAADF-STEM) images of $KPb_2Cl_5:Nd^{3+}$ nanocrystals were acquired using a FEI ThemIS 60−300 STEM/TEM (ThermoFisher Scientific, US) operated at 300 kV at the LBNL National Center for Electron Microscopy. Energy Dispersive

Spectroscopy (EDS) mapping was performed using a Bruker Super-X Quad windowless detector with a solid angle of 0.7 sr. TEM samples were prepared by drop-casting nanoparticles dispersed in hexane onto a copper TEM grid with an ultra-thin graphene layer (Ted Pella). The TEM imaging of $KPb_2Cl_5:Nd^{3+}$ and $NaYF_4:Nd^{3+}$ nanocrystals was also performed with a transmission electron microscope (JEOL 2100F) operated at an accelerating voltage of 200 kV. The Nd to Pb atomic fraction on the EDS map was calculated by removing the background signal contribution and integrating the Nd and Pb X-ray signals for each nanocrystal region. To avoid potential X-ray signal overlap, only Nd-L and Pb-L were used to calculate the atomic fractions of Nd and Pb. The molar percentage of $Nd^{3+}$ dopants in $KPb_2Cl_5$ nanocrystals was measured by elemental analysis using inductively coupled plasma optical emission spectroscopy (Varian ICP-OES 720 Series)[1].

### 1.5 Optical characterization

Photon avalanche emission of $KPb_2Cl_5:Nd^{3+}$ nanocrystals was measured in NP films prepared by drop-casting chloroform dispersions with NPs onto glass coverslips (0.13 mm thickness) or silicon substrates. Films were characterized using a custom-built confocal inverted microscope (see principal scheme below) equipped with Lake Shore Janis ST-500 cryostat. The sample was excited by a 1064 nm continuous-wave laser source (Opus 1064 3000, Laser Quantum) coupled through a 950 nm short-pass dichroic mirror and a long working distance air objective (60x 0.7NA, Nikon). Emission spectra were collected by the same objective, spectrally filtered using a 950 short-pass filter, and imaged onto an EMCCD camera (Andor iXon Ultra 897) equipped spectrometer (Acton Research Corp., SpectraPro-300i). Absolute intensities of $Nd^{3+}$ emission lines at 810 nm or 880 nm were measured using an avalanche photodiode and 809/81 nm or 935/170 nm band-pass filters, respectively. For power dependence measurements, a continuously variable, reflective neutral density filter wheel (Thorlabs) was inserted in the laser beam path for coarse power selection, and fine power steps were obtained by motorized rotation of a half-wave plate coupled with a Glan-Taylor prism (Thorlabs), power selection was synchronized and automated with the collection system. Powers were simultaneously recorded by a Thorlabs power meter from a glass coverslip to reflect ~10% of the incoming flux. Average excitation power densities were calculated using measured laser powers and the $1/e^2$ area for the employed excitation wavelength and microscope objective. Photoswitching experiments were done by co-aligning Ti:Sapphire laser (Chameleon Ultra II, Coherent) onto the sample. All power-dependent data is presented and analyzed without post-acquisition treatment (e.g., background subtraction).

Temporally-resolved luminescence data was collected with APD coupled to a time-correlated single photon counter (TCSPC, HydraHarp 400), which was used to tag photon arrival times of luminescence signal with respect to the laser shutoff trigger event. An optical chopper was used to vary the excitation frequency from 20 to 1000 Hz (variable duty cycle blade: MC1F10A, Thorlabs) and from 120 to 6000 Hz (50% duty cycle blade: MC1F60, Thorlabs).

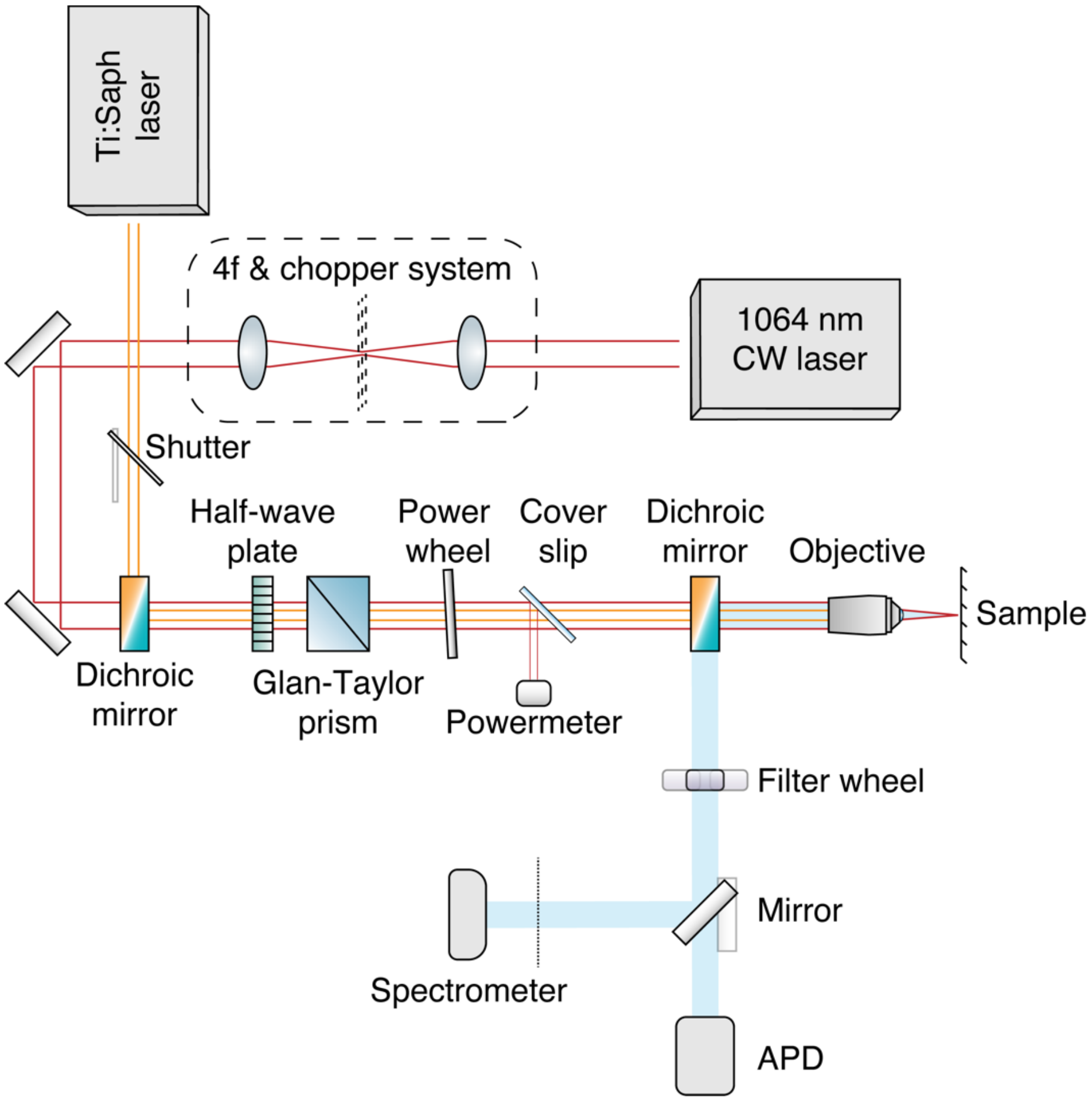

Principle arrangement of the custom-built confocal laser scanning microscope, equipped with 1064 nm CW laser, tunable-wavelength Ti:Saph laser, automatized coarse and fine power adjustment by power wheel and (half-wave plate + Glan-Taylor prism), respectively, and spectrometer or APD baser photoluminescence detection.

### 1.6 Analysis of luminescence hysteresis

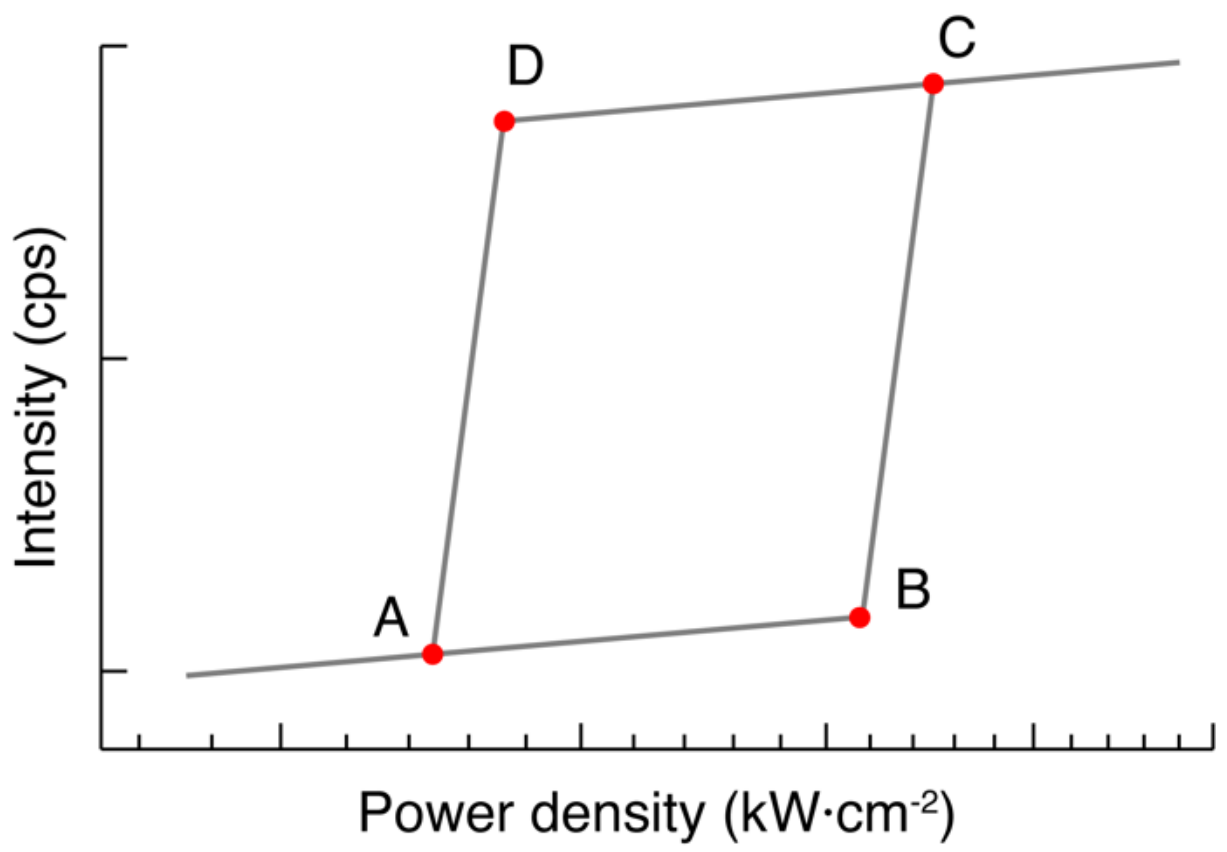

A schematic representation of an intensity (I) vs power density (P) graph, with key transition points of the hysteresis curve indicated by alphabetical indices A, B, C, and D.

To analyze photoluminescence hysteresis of $KPb_2Cl_5:Nd^{3+}$ nanocrystals, we have estimated the switch-on threshold, hysteresis width, and switch-on contrast from the power-dependent photoluminescence plots (schematically represented as *I* vs *P* in the above figure) as follows:

$$Switch-on\ theshold = P_B \tag{S1}$$

$$Hysteresis\ width = \frac{P_C - P_B}{2} - \frac{P_A - P_D}{2} \tag{S2}$$

$$Switch-on\ contrast = 10 * log_{10}\left[\frac{(I_C - I_B)}{I_{noise}}\right] \tag{S3}$$

Here, $I_{noise}$ = 82 cps, the average background noise intensity attainable with our setup.

## 2. Structural characterization of $KPb_2Cl_5:Nd^{3+}$ ANPs

To characterize $KPb_2Cl_5:Nd^{3+}$ ANPs, we performed a detailed TEM/STEM study on the as-synthesized samples from two different batches (Supplementary Figure S1). The as-synthesized $KPb_2Cl_5:Nd^{3+}$ ANPs exhibit prolate morphology with an average length of 61±13 nm and 63±10 nm, measured for the two batches of nanocrystals. The larger size and lower surface-area-to-volume ratio of these nanocrystals minimize the non-radiative quenching of $Nd^{3+}$ dopants by surface vibrations. We further performed STEM-EDS analysis on two isolated $KPb_2Cl_5:Nd^{3+}$ ANPs (Supplementary Figure S2) and observed a homogeneous distribution of elements. From the corresponding EDS spectra, we calculate the $Nd^{3+}$ doping concentration ([$Nd^{3+}$] per Pb site) to be 12 ± 2 % and 17 ± 3 % in the two ANPs, which is in good agreement with the average [$Nd^{3+}$] analyzed via ICP-OES.

We note that $KPb_2Cl_5$ crystals have two crystallographically distinct $Pb^{2+}$ sites with coordination numbers (CN) = 7 for Pb(1) and 9 for Pb(2) [3,4]. Since $Ln^{3+}$ incorporation is favored in sites with greater CN [5], $Ln^{3+}$ ions preferentially substitute Pb(2) [3]. Subsequently, when $Ln^{3+}$ ions enter the $KPb_2Cl_5$ matrix, a $K^+$ vacancy is formed to compensate for the charge imbalance[3,6].

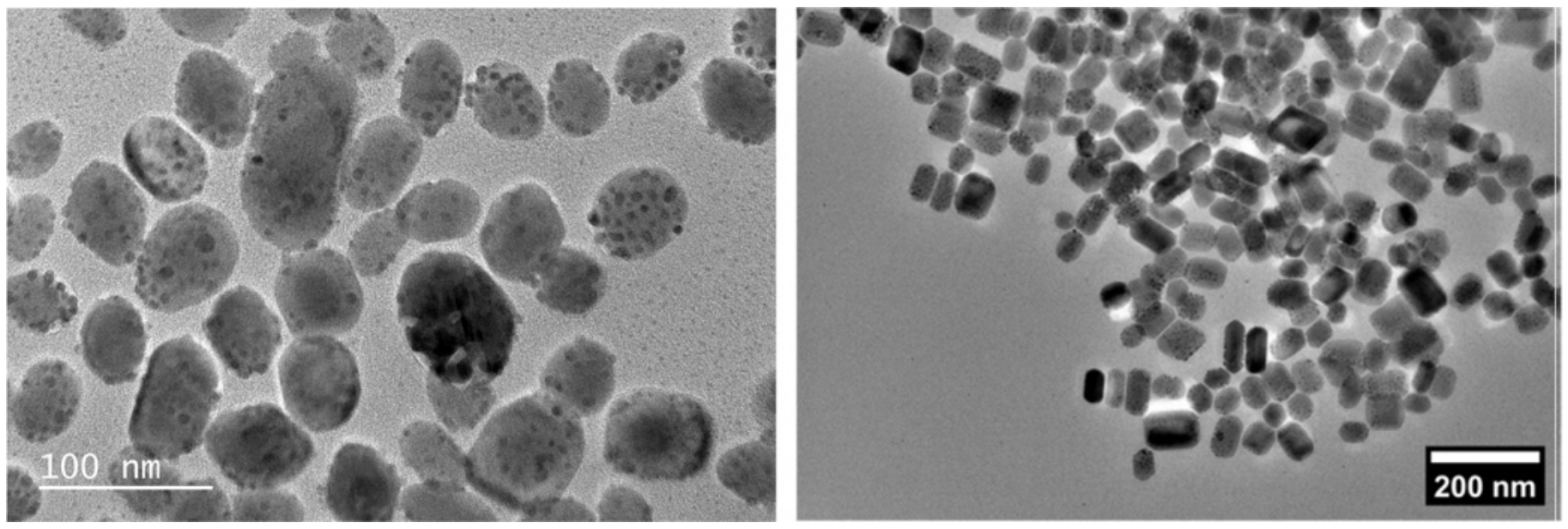

**Supplementary Figure S1.** Representative TEM images of the bistable $KPb_2Cl_5:Nd^{3+}$ photon avalanching nanoparticles (ANPs) from two different synthesis batches. We note that exposure to e-beam induces degradation of the nanocrystals, seen as the appearance of the darker contrast $PbCl_2$ spots, consistent with other reports[7–9]. Importantly, these $PbCl_2$ accumulations result from nanocrystals being mildly sensitive to the environment (e.g., moisture or e-beam) during long-term storage but do not impact the performance of as-synthesized or nondegraded $KPb_2Cl_5:Nd^{3+}$ ANPs.

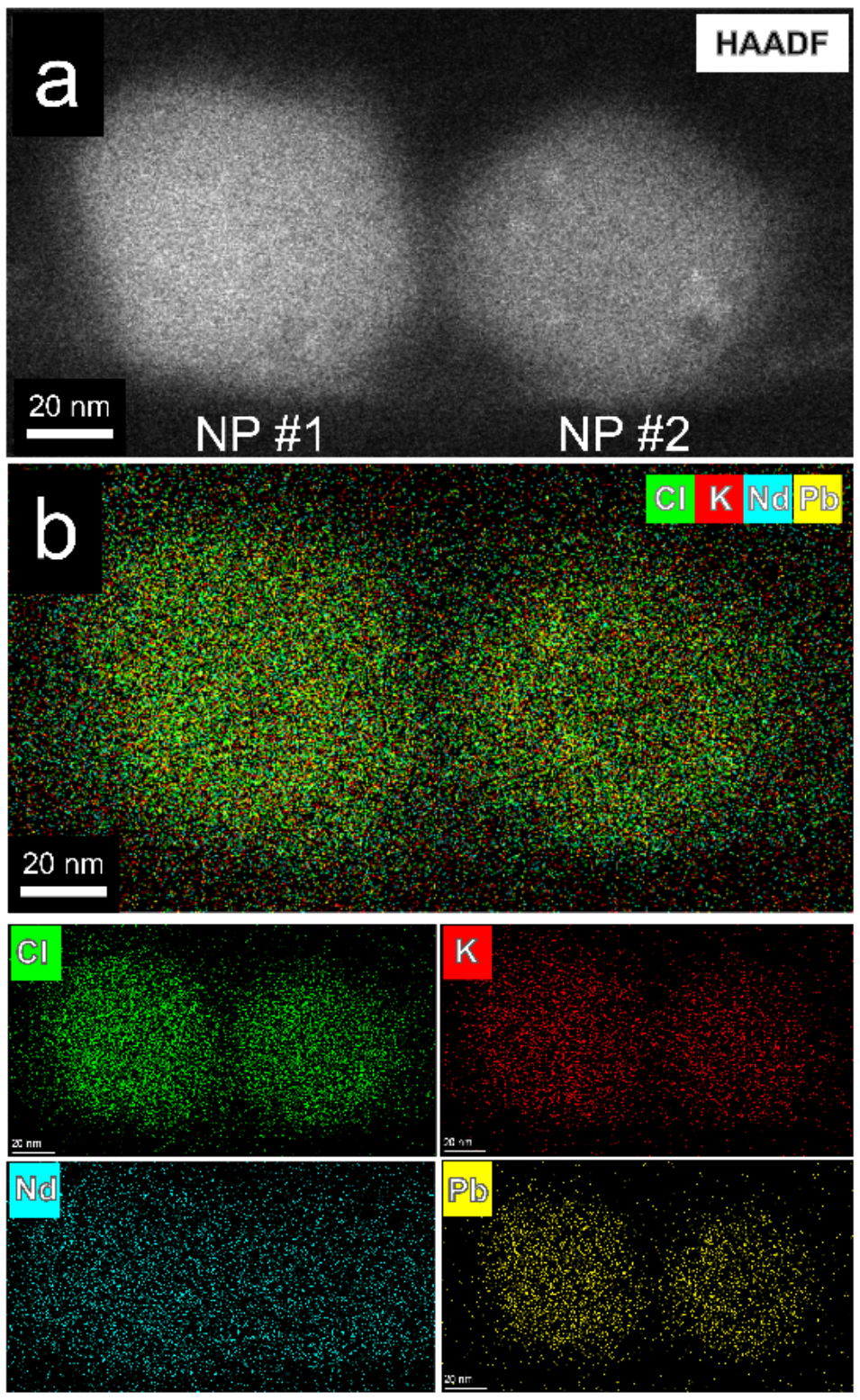

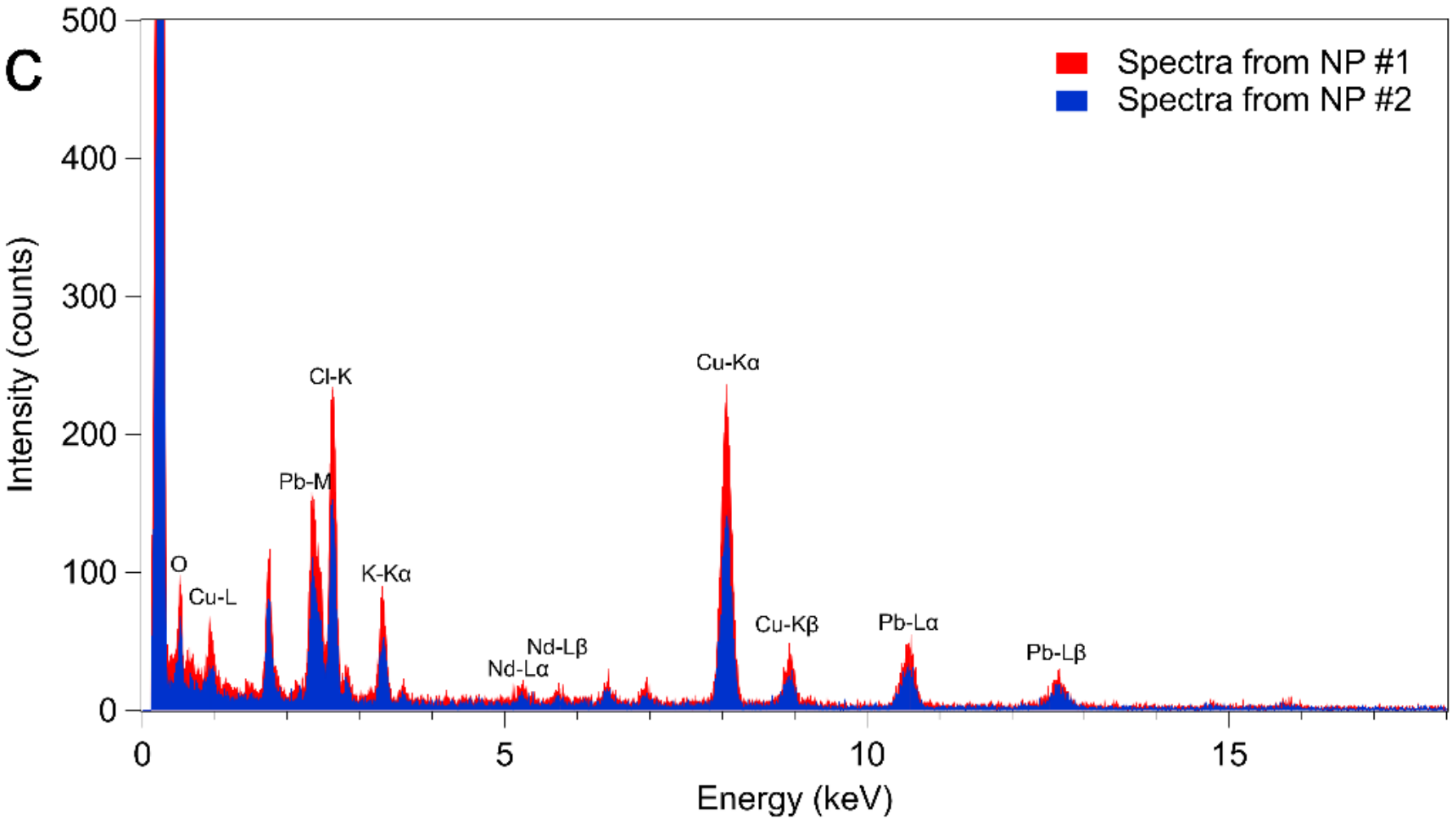

| Nanoparticle | Nd Atomic Fraction (%) | Pb Atomic Fraction (%) |
|---|---|---|
| #1 | 12 ± 2 | 88 ± 14 |
| #2 | 17± 3 | 83 ± 13 |

**Supplementary Figure S2**. **A** – HAADF-STEM image of two $KPb_2Cl_5:Nd^{3+}$ nanocrystals analyzed via STEM-EDS for elemental composition. Analyzed $KPb_2Cl_5:Nd^{3+}$ nanocrystals are referred to as NP #1 (left nanocrystal) and NP #2 (right nanocrystal). **B** – Corresponding STEM-EDS elemental mapping of merged and individual Cl-K, K-K, Nd-L, and Pb-L channels. **C** – EDS spectra of the two analyzed nanocrystals, with the calculated Nd to Pb atomic fractions of 12% and 17% in NP #1 and NP #2, respectively. This result is in good agreement with the average [$Nd^{3+}$], analyzed via ICP-OES[1].

## 3. MPR rate vs host phonon energy and temperature

The scaling of MPR rate ($W$) was determined using[10]:

$$W = W_0 e^{-(\beta \Delta E)} \left(1 - exp\left[\frac{\hbar\omega}{k_B T}\right]\right)^{-p} \quad \text{(S4)}$$

where $W_0$ and $\beta$ are host-dependent parameters, $\Delta E$ energy gap between adjacent states, $\hbar\omega$ host phonon energy, $k_B$ Boltzmann constant, $T$ temperature, and $p$ number of phonons involved in bridging the energy gap:

$$p = \frac{\Delta E}{\hbar\omega} \quad \text{(S5)}$$

**Supplementary Table S1**. Host-dependent phonon relaxation parameters and energy gap values between first excited and ground states of $Nd^{3+}$ ions in $NaYF_4$ and $KPb_2Cl_5$ hosts.

| Parameter [units] | $KPb_2Cl_5$ | $NaYF_4$ |
|---|---|---|
| Phonon cutoff energy $\hbar\omega$ [$cm^{-1}$] | 250 ref:1,6 | 450 ref:11 |
| $W_0$ rate constant [$s^{-1}$] | $0.5 \cdot 10^7$ | $1.0 \cdot 10^7$ |
| $\beta$ MPR rate constant [cm] | $9.2 \cdot 10^{-3}$ | $2.0 \cdot 10^{-3}$ |
| $\Delta E$ energy gap ($^4I_{11/2}$–$^4I_{9/2}$) [$cm^{-1}$] | 1900 | 1828 |

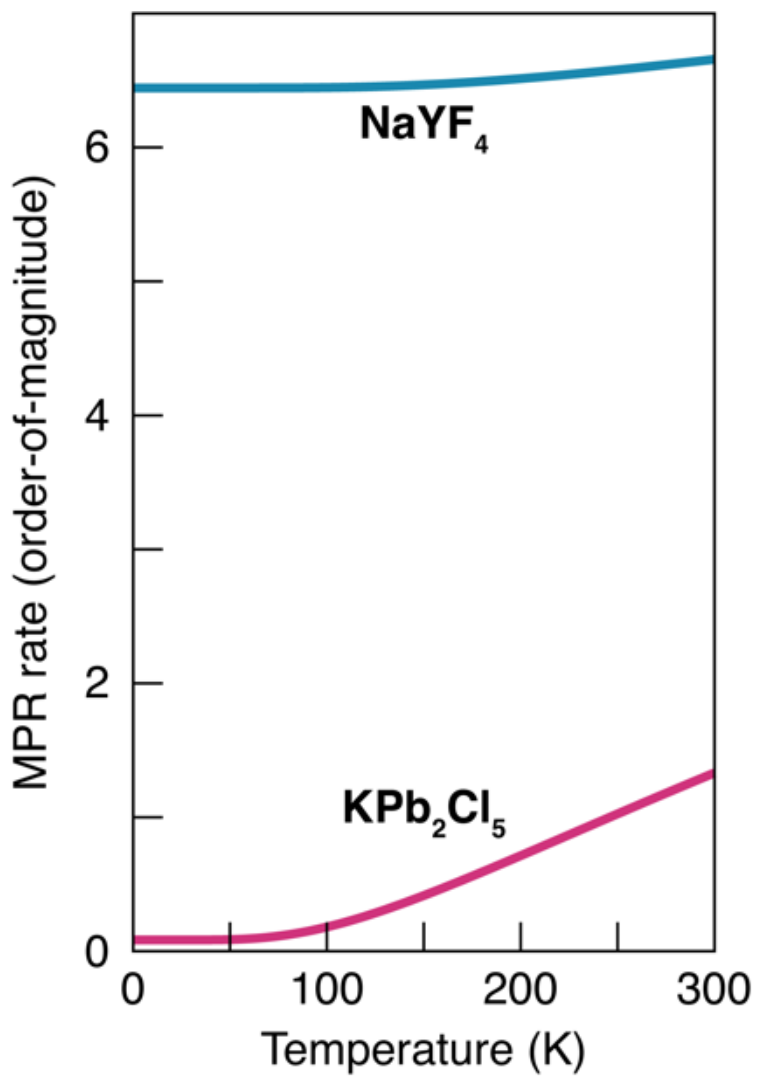

**Supplementary Figure S3.** Calculated temperature-dependent scaling of multiphonon relaxation (MPR) rates between the first excited and ground state ($^4I_{11/2} \rightarrow {}^4I_{9/2}$) of $Nd^{3+}$ ions in $NaYF_4$ and $KPb_2Cl_5$ hosts.

## 4. Robustness of IOB to measurement conditions

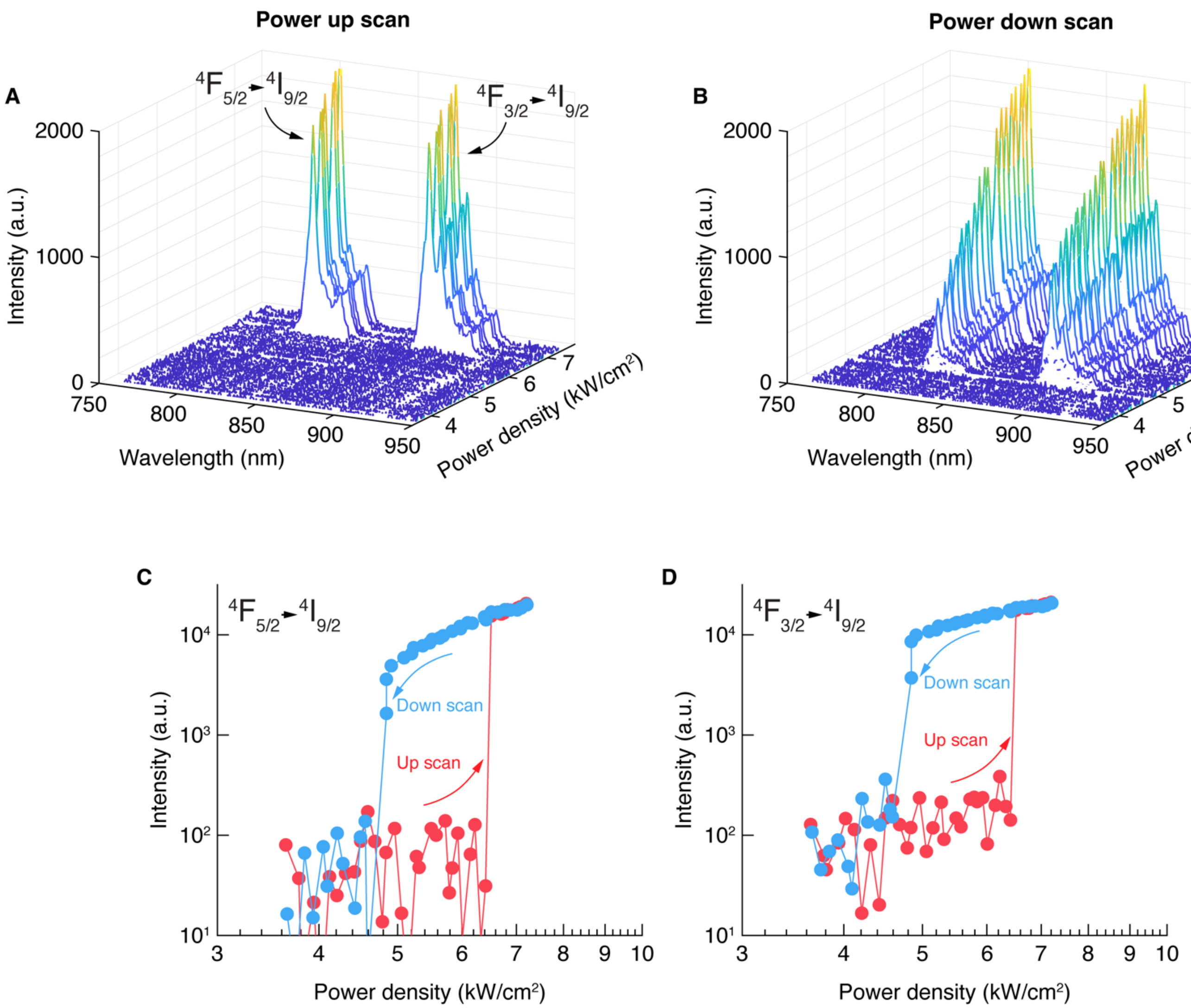

**Supplementary Figure S4. A, B** – Pump (1064 nm) power dependence of $Nd^{3+}$ emission in $KPb_2Cl_5:Nd^{3+}$ ANPs at 77 K registered with a spectrometer, instead of an APD as in all other power-dependent measurements. Spectra of ANPs can be observed at lower pump powers in the down scan (**B**) compared to the up scan (**A**). **C, D** – Integrated $Nd^{3+}$ emission intensity in $KPb_2Cl_5:Nd^{3+}$ ANPs at 77 K against pump power, corresponding to $^4F_{5/2} \rightarrow {}^4I_{9/2}$ (integration range 800-830 nm, **C**) and $^4F_{3/2} \rightarrow {}^4I_{9/2}$ (integration range 870-900 nm, **D**) radiative transitions. Red data points represent scanning with increasing powers (up scan) and blue with decreasing (down scan).

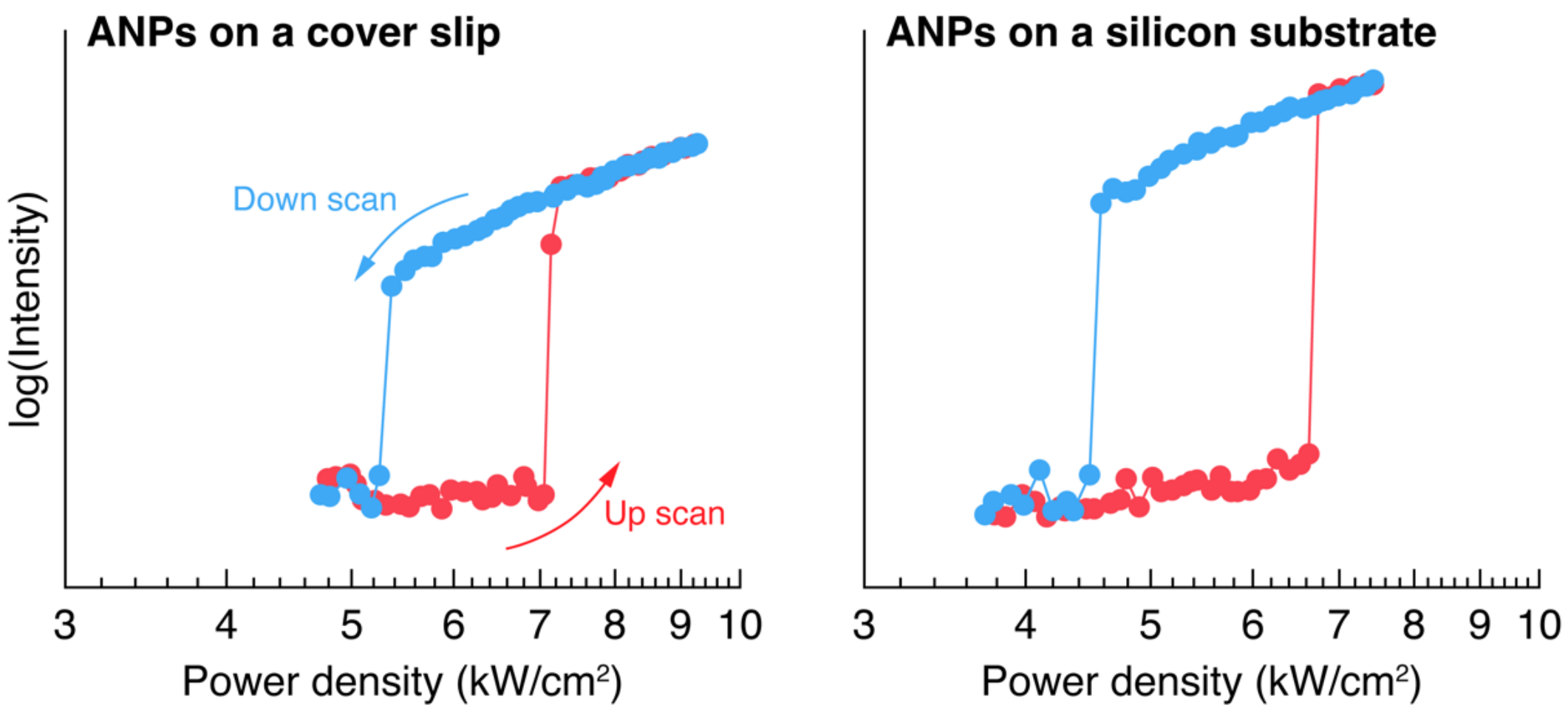

**Supplementary Figure S5.** Pump (1064 nm) power dependence of $Nd^{3+}$ emission (810 nm, $^4F_{5/2} \rightarrow {}^4I_{9/2}$) in $KPb_2Cl_5:Nd^{3+}$ ANPs at 77 K. ANPs were deposited either on a coverslip (left) or on a silicon substrate (right). Red data points represent scanning with increasing powers (up scan) and blue with decreasing (down scan).

## 5. Measuring IOB in $KPb_2Cl_5:Nd^{3+}$ ANPs using different photodiode dwell times

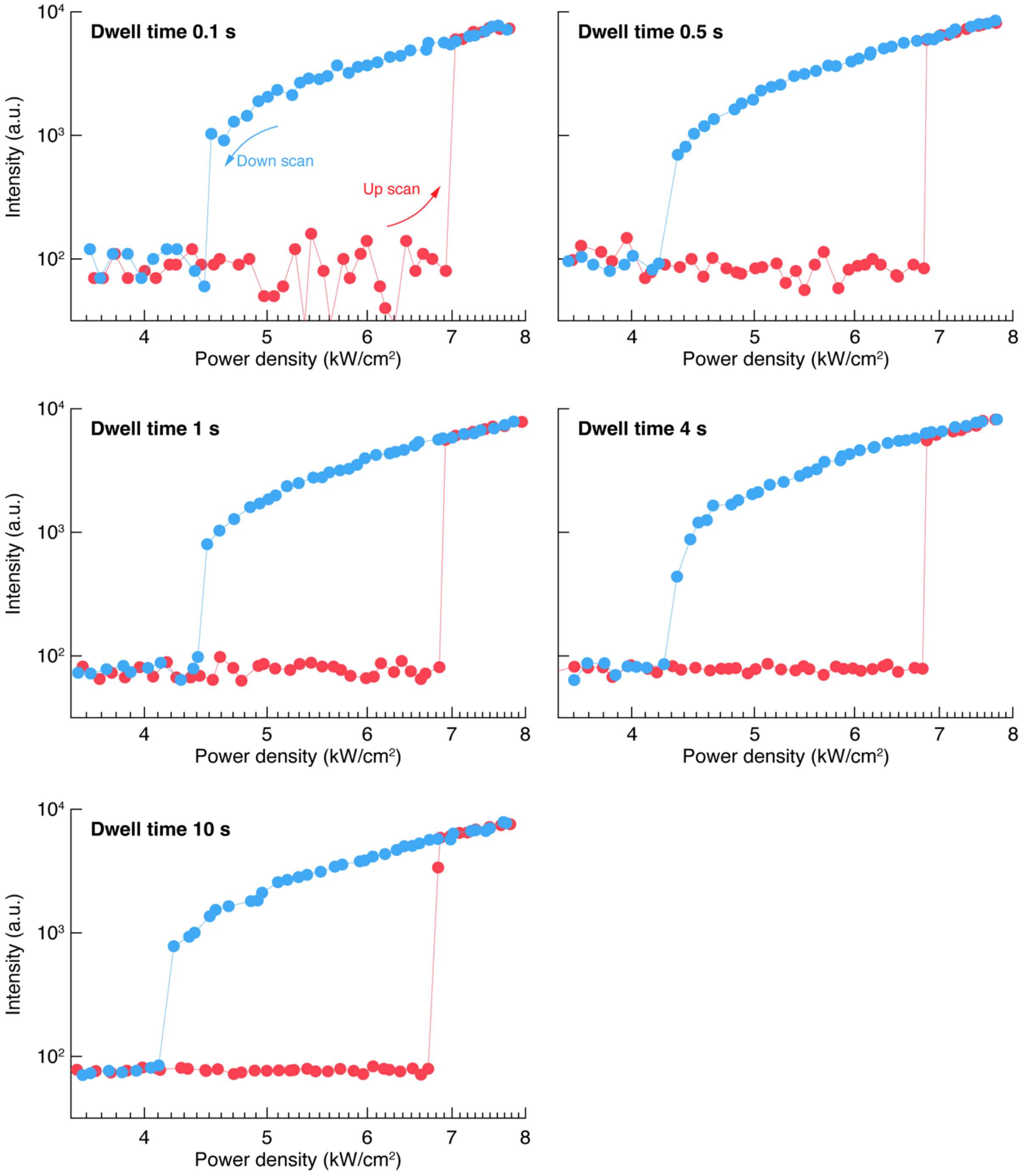

**Supplementary Figure S6.** Pump (1064 nm) power dependence of $Nd^{3+}$ emission (810 nm, $^4F_{5/2} \rightarrow {}^4I_{9/2}$) in $KPb_2Cl_5:Nd^{3+}$ ANPs at 77 K, using different signal dwell time per set power value. Red data points represent scanning with increasing powers (up scan) and blue with decreasing (down scan). Here, dwell time corresponds to the time taken by the APD to measure luminescence intensity in counts-per-second (cps) at a given excitation power density. Longer dwell times are equivalent to signal averaging and reduce uncertainty in the determined luminescence intensity. A signal collection delay of 0.5 s was implemented to ensure ANPs reached a steady state before each photoluminescence measurement.

## 6. IOB in $KPb_2Cl_5:Nd^{3+}$ ANPs: reproducibility and repeatability

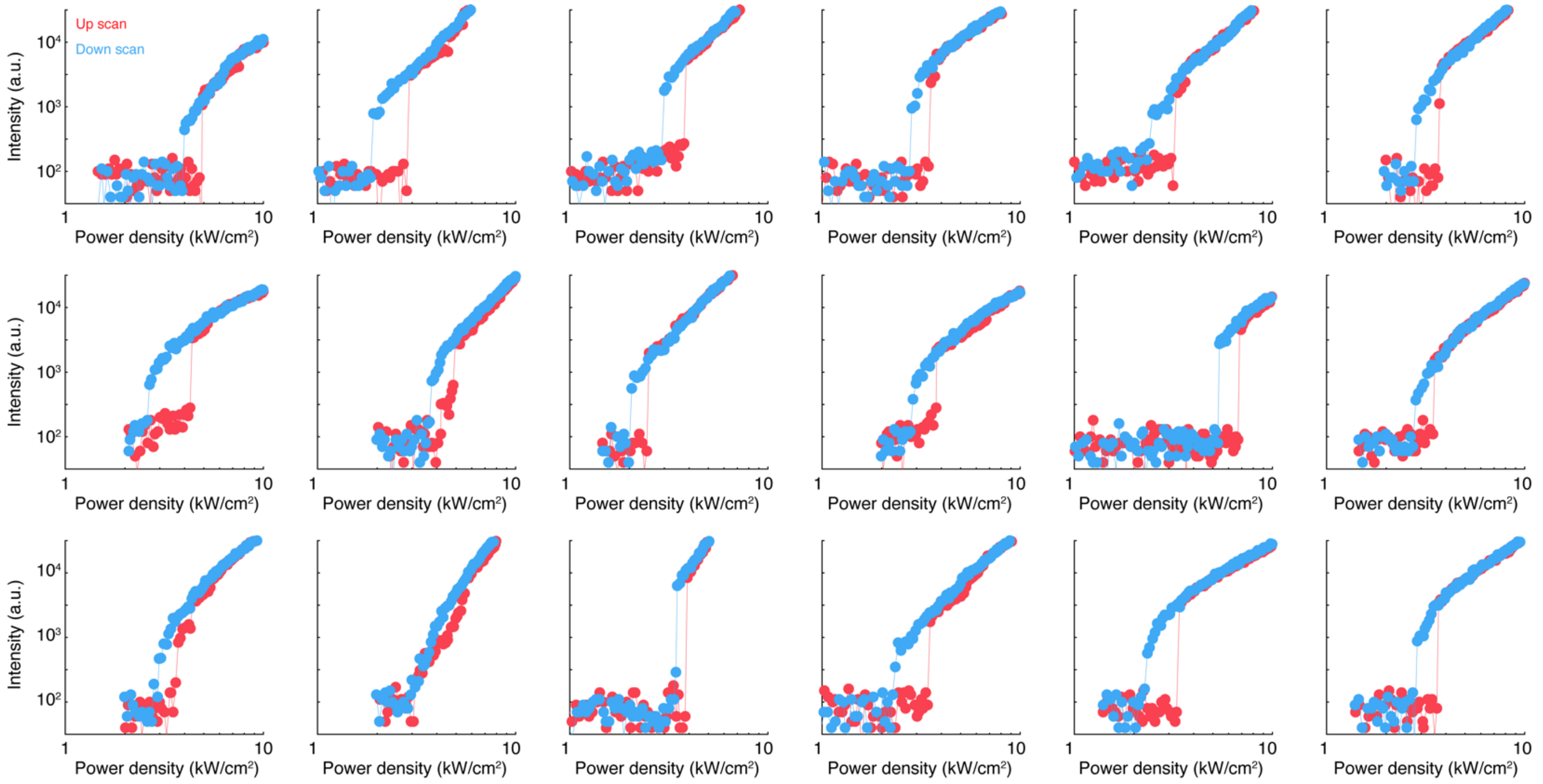

**Supplementary Figure S7.** Representative pump (1064 nm) power dependence curves of $Nd^{3+}$ emission (810 nm, $^4F_{5/2} \rightarrow {}^4I_{9/2}$) in $KPb_2Cl_5:Nd^{3+}$ ANPs at 77 K. Data were taken for multiple spots of the film sample. Red data points represent scanning with increasing powers (up scan) and blue with decreasing (down scan).

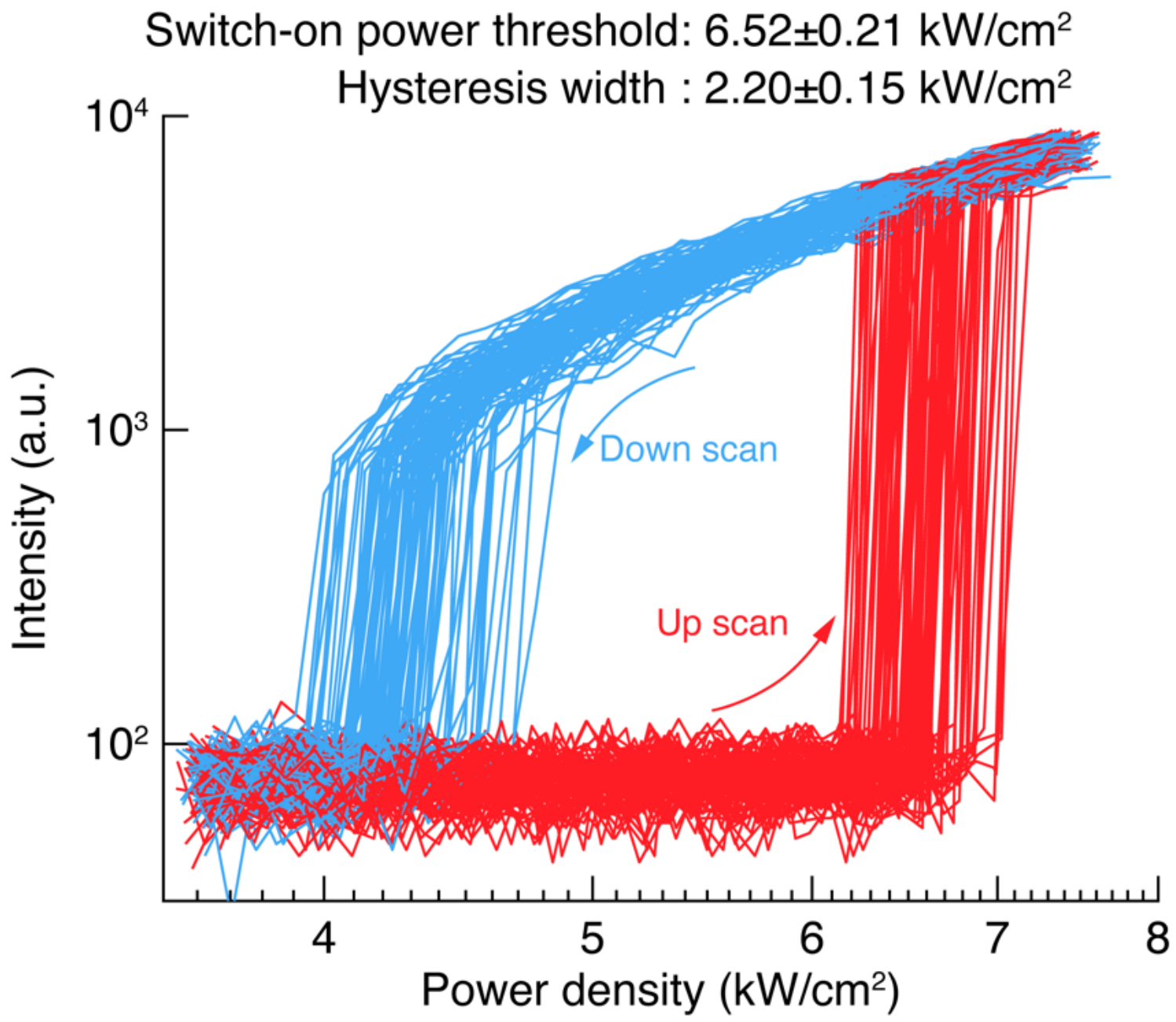

**Supplementary Figure S8.** Pump (1064 nm) power dependence of $Nd^{3+}$ emission (810 nm, $^4F_{5/2} \rightarrow {}^4I_{9/2}$) in $KPb_2Cl_5$: $Nd^{3+}$ ANPs at 77 K measured consecutively across 100 cycles over 12 hours. Red lines represent scanning with increasing powers (up scan) and blue with decreasing powers (down scan). Average ± standard deviation values of the switch-on threshold and hysteresis width are shown above the graph.

**7. Supplementary Table S2. Comparison of different materials and systems that show optical bistability.** Switch-on contrast and threshold values are best estimates for works that do not provide them explicitly. [a] Average observed contrast. Contrast calculated as the intensity difference between bright and dark states divided by the noise intensity. For literature values, we report maximum contrast represented. [b] Contrast as calculated by the authors. [c] Data to assess the size is not provided. [d] The origin of bistability is either uncertain or fits better the explanation given by ref. 12. [e] Qualitative estimate since intensity values are not provided.

| Material/system | Scale (size) | Excitation wavelength | Observable | Temperature | Switching contrast | Switching threshold | Origin of bistability | Ref. |
|---|---|---|---|---|---|---|---|---|
| Silver-coated CdS | Nano (15 nm) | 514.5 nm | Scatter/transmission | RT | 1.4 | 52 kW/cm$^2$ (~650 mW) | Non-thermal | 13 |
| Amorphous silicon Mie resonator | Nano (155 x 142 nm) | 785 nm | Scatter/transmission | RT | 10 | 360 kW/cm$^2$ | Thermal | 14 |
| Silicon nanowire-gold nanorod hybrid | Nano (15 nm x 16 μm) | 1310 nm | Scatter/transmission | RT | 14 | 11.7 μW | Non-thermal | 15 |
| $KPb_2Cl_5:Nd^{3+}$ | Nano (61 nm) | 1064 nm | Photoluminescence | 77-150 K | 218 (max.) 43 (mean)[a] | 5.2 kW/cm$^2$ (64 μW) | Non-thermal | This work |
| $Yb^{3+}$, $Er^{3+}$-co-doped phosphate glass | Micro (50 μm) | 1550 nm | Photoluminescence | RT | 21[b] | 8.5 mW | Thermal | 16 |
| InP/InGaAsP photonic crystal | Micro (4 x 0.3 x 0.15 μm) | 1550 nm | Scatter/transmission | RT | 100 | 25 nW | Non-thermal | 17 |
| $ZrO_2:Yb^{3+},Tm^{3+}$ | Bulk nanocrystalline[c] | 973 nm | Photoluminescence | RT | 2 | >200 mW | Thermal | 18 |
| $LiNdP_4O_{12}$ | Bulk nanocrystalline[c] | 808 nm | Photoluminescence | RT | 3.2 | >100 mW | Tentatively thermal[d] | 19 |
| $Cs_3Yb_2Br_9$ | Bulk | 944 nm | Photoluminescence | 5-15 K | High vs noise[e] | 2.27 kW/cm$^2$ | Tentatively thermal[d] | 20 |
| $NdPO_4:Yb^{3+}$ | Bulk | 808 nm | Photoluminescence | 15-200 K | Low vs noise[e] | 0.5 kW/cm$^2$ | Thermal | 21 |
| $CsCdBr_3:Yb^{3+}$, $Er^{3+}$ | Bulk | 943 nm | Photoluminescence | 6-25 K | High vs noise[e] | 4 kW/cm$^2$ | Thermal | 22 |
| $CsCdBr_3:Yb^{3+}$ | Bulk | 944 nm | Photoluminescence | 14-35 K | High vs noise[e] | 4 kW/cm$^2$ | Tentatively thermal[d] | 23 |
| $Y_2O_3$ | Bulk | 980 nm | Photoinduced blackbody radiation | RT | 25 | 22 W | Non-thermal | 24 |
| $Yb^{3+}$, $Tm^{3+}$-co-doped glass | Bulk | 960 nm | Photoluminescence | RT | High vs noise[e] | 370 mW | Tentatively thermal[d] | 25 |
| $Y_2WO_6:Yb^{3+},Tm^{3+}$ | Bulk | 980 nm | Photoluminescence | RT | 1.5 | 453 mW | Tentatively thermal[d] | 26 |
| $Er^{3+}$-doped fiber | Bulk | 980 nm | Photoluminescence | RT | 2500 | 72 mW | - | 27 |
| $Yb^{3+}$-doped high silica glass | Bulk | 980 nm | Photoluminescence | RT | 1.3 | >2.5 W | Tentatively thermal[d] | 28 |
| InSb cavity | Bulk | 5497 nm | Scatter/transmission | 77 K | 1.2 | 26 mW | Non-thermal | 29 |

## 8. IOB in $KPb_2Cl_5:Nd^{3+}$ ANPs against $Nd^{3+}$ doping concentration

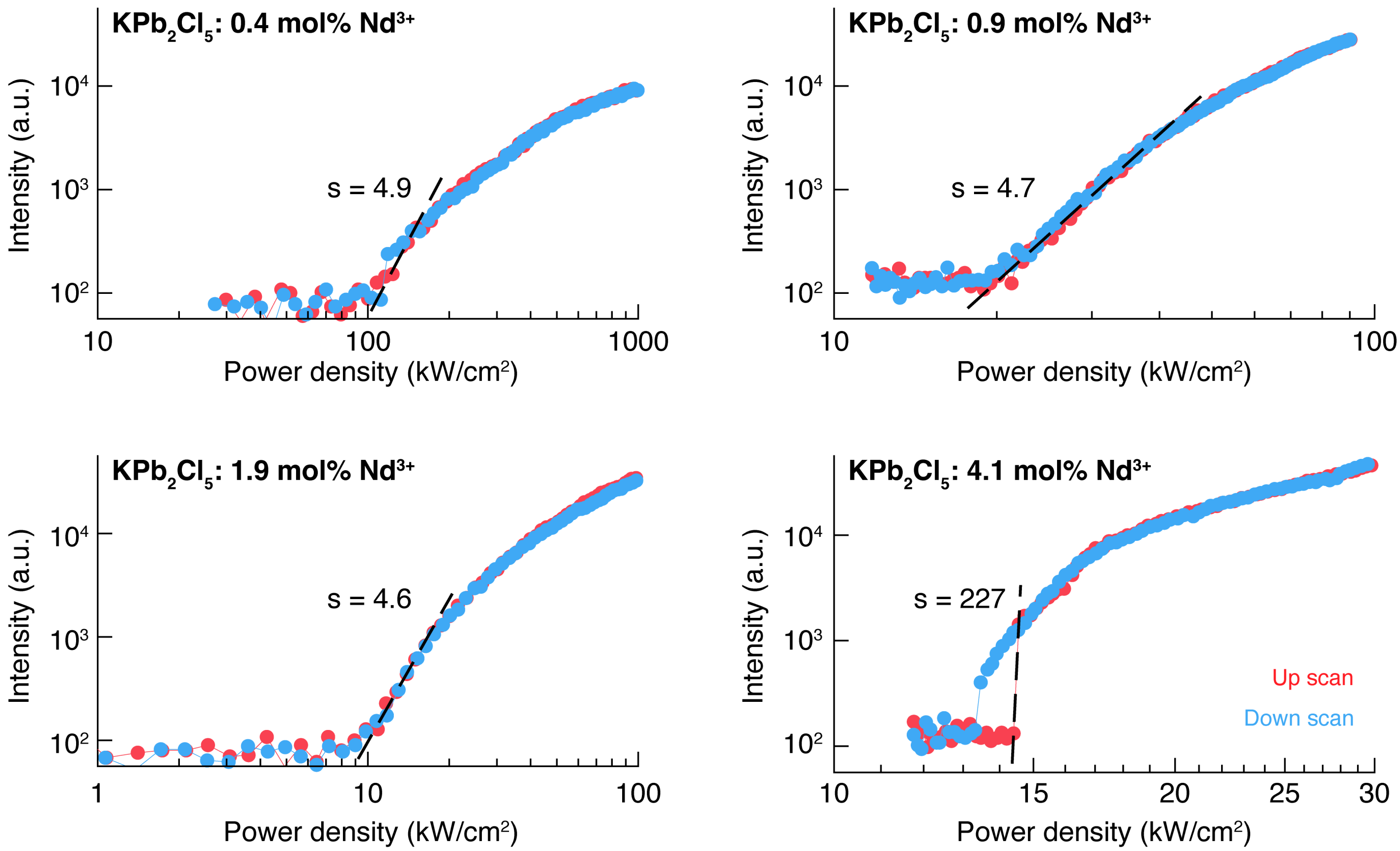

**Supplementary Figure S9.** Representative pump (1064 nm) power dependence of $Nd^{3+}$ emission (810 nm, $^4F_{5/2} \rightarrow {}^4I_{9/2}$) in $KPb_2Cl_5$: x mol% $Nd^{3+}$ ANPs at 77 K. Red data points represent scanning with increasing powers (up scan) and blue with decreasing (down scan).

## 9. Power dependence of $NaYF_4$:$Nd^{3+}$ nanocrystals at cryogenic temperatures

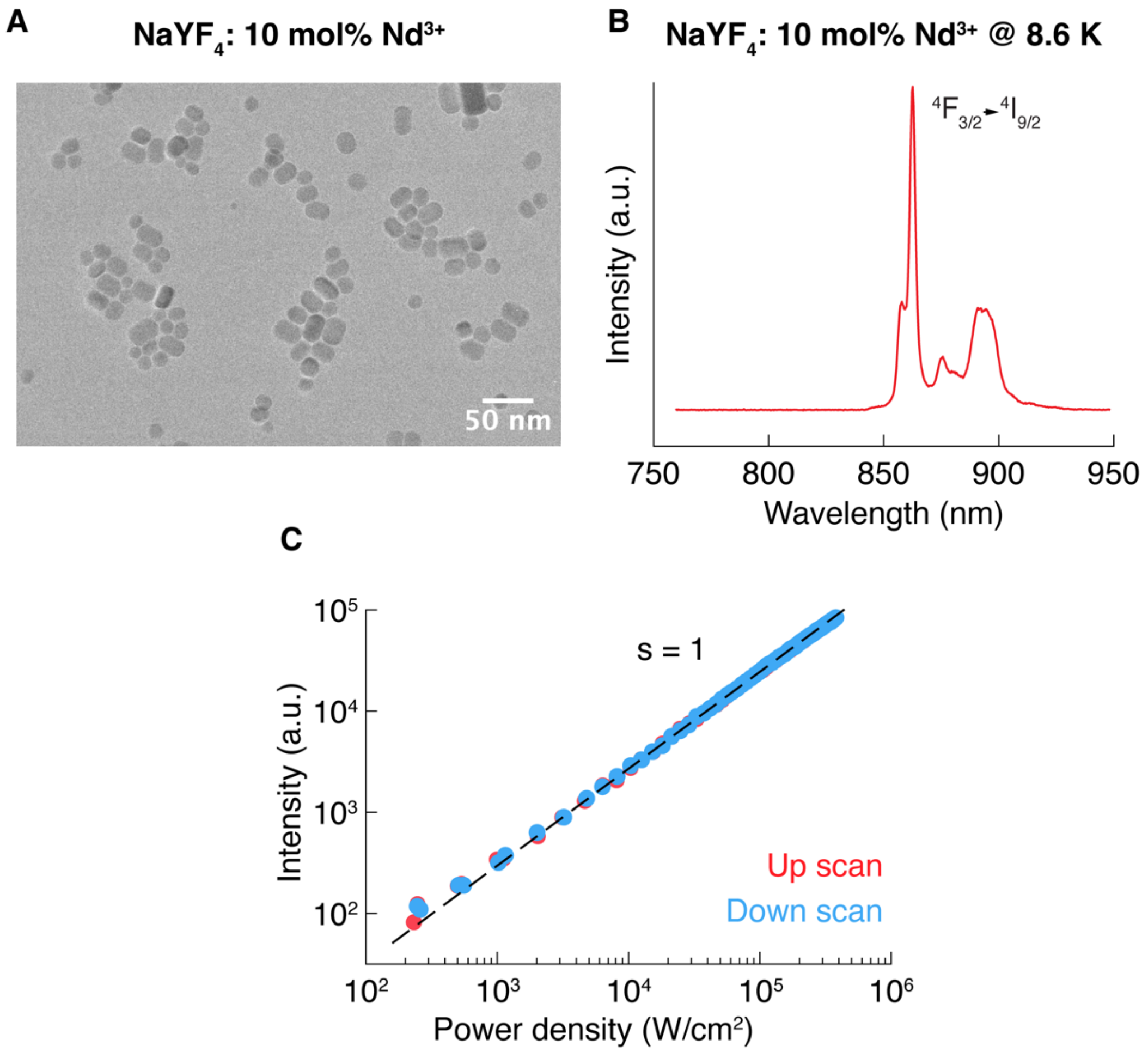

**Supplementary Figure S10. A** – Representative TEM image $NaYF_4$: 10 mol% $Nd^{3+}$ NPs. **B** – Emission spectrum of $NaYF_4$: 10 mol% $Nd^{3+}$ NPs at 8.6 K under 1064 nm excitation. **C** – Pump (1064 nm) power dependence of $Nd^{3+}$ emission (880 nm, $^4F_{3/2} \rightarrow {}^4I_{9/2}$) in $NaYF_4$:$Nd^{3+}$ NPs at 8.6 K. Red data points represent scanning with increasing powers (up scan) and blue with decreasing (down scan).

## 10. Supplementary Note S1. Rate constants and rate equation model

*Theoretical description of the system.* Since the intensity of a given radiative transition is proportional to the product of the population of the emitting state and the microscopic rate constant for the transition, we used a set of coupled differential equations to describe the instantaneous populations of the $Ln^{3+}$ manifolds[30]. In this population balance model, the population $N_i$ (in $nm^{-3}$) of a lanthanide $4f^N$ manifold $i$ over time is determined by the incoming and outgoing rates of electric dipole (ED) and magnetic dipole (MD) radiative transitions, nonradiative multiphonon relaxation (MPR) and energy transfer (ET):

$$\frac{dN_i}{dt} = \sum_j \left(N_j A_{ji}^{ED} - N_i A_{ij}^{ED}\right) + \sum_j \left(N_j A_{ji}^{MD} - N_i A_{ij}^{MD}\right) + \left(N_{i+1} W_{i+1,i}^{NR} - N_i W_{i,i-1}^{NR}\right) + \sum_{ij,kl} \left(N_j N_l P_{ji,kl}^{ET} - N_i N_k P_{ij,lk}^{ET}\right) \quad \text{(S6)}$$

$A_{ij}^{ED}$ and $A_{ij}^{MD}$ ($s^{-1}$) are the Einstein coefficients for ED and MD radiative transitions from manifold $i$ to $j$. $W_{i,i-1}^{NR}$($s^{-1}$) is the nonradiative MPR rate from the manifold $i$ to the manifold immediately below $i$. $P_{ij,kl}^{ET}$ ($nm^3 \cdot s^{-1}$) is the microscopic energy transfer parameter for the transfer of energy for the donor $i$ to $j$ transition and the acceptor $k$ to $l$ transition.

Transition probabilities for individual lanthanide ions are calculated using the same theoretical framework as the comprehensive rate equation models described in our previous work[30,31] (see Ref. 32 for the most detailed description). The transitions in trivalent lanthanide ions are $4f^N$ manifold-to-manifold transitions, including single-ion transitions (non-radiative multi-phonon relaxation, electric dipole absorption/emission, and magnetic dipole absorption/emission) and two-ion interactions (dipole-dipole energy transfer). The individual theories and equations behind these transition probabilities are discussed below. Representative rate constants comparing $KPb_2Cl_5:Nd^{3+}$ and $NaYF_4:Nd^{3+}$ are given in Supplementary Table S3.

**Multiphonon relaxation (MPR) transition probabilities**, $W_i^{MPR}$, from an initial lanthanide manifold $i$ to the level below $i$ are calculated using Van Dijk's modified Energy Gap Law[33]:

$$W_i^{NR} = W_0^{MPR} e^{-\alpha(\Delta E - 2\hbar\omega)} \quad \text{(S7)}$$

where $\Delta E$ is the energy gap to the lower manifold and $\hbar\omega$ is the energy of the cutoff phonon energy of the host matrix. $W_0^{MPR}$ and $\alpha$ are MPR constants that are consistent across transitions in a given host matrix. $W_0^{MPR}$ is the MPR rate when $\Delta E = 2\hbar\omega_{max}$,

**Electric dipole (ED) radiation and absorption probabilities** are calculated using Judd-Ofelt (JO) theory[32,34]. The transition probability (in $s^{-1}$) for electric dipole radiative relaxation, $A_{ij}^{ED}$, is determined using the three JO parameters ($\Omega_\lambda$) and the doubly reduced matrix elements of the unit tensor operator $U^{(\lambda)}$ of rank $\lambda$, where $\lambda =$ 2,4, or 6:

$$A_{ij}^{ED} = \frac{64\pi^4 e^2 \tilde{\upsilon}^3}{3h(2J+1)}[\frac{n(n^2+2)^2}{9}]S_{ij}^{ED} \tag{S8}$$

$$S_{ij}^{ED} = \sum_{\lambda=2,4,6} \Omega_\lambda | < [SL]J_i||U^{(\lambda)}||[SL]J_j > |^2 \tag{S9}$$

Here, $\tilde{\upsilon}$ is the energy of the transition in wavenumbers, $h$ is Planck's constant, $e$ is the elementary charge, $n$ is the index of refraction of the material, $J$ is the total angular momentum of the initial state, and $S_{ij}^{ED}$ is the electric dipole line strength, using the Gaussian unit system. JO parameters used in this work are shown in Supplementary Table S3.

**Supplementary Table S3.** Judd-Ofelt parameters and reduced matrix elements.

| Parameter | $Nd^{3+}$ in $KPb_2Cl_5$ |
|---|---|
| $\boldsymbol{\Omega_2}$ ($10^{-20}$ $cm^2$) | 11 |
| $\boldsymbol{\Omega_4}$ | 8.01 |
| $\boldsymbol{\Omega_6}$ | 6.19 |
| Source, $\boldsymbol{\Omega_\lambda}$ | Isaenko et al.[35] |
| Source, $\lvert<i\lvert U_\lambda \rvert j>\rvert^2$ | Kaminskii et al.[10] |

For ED absorption, $A_{ij} = \psi\sigma_{ij}$, where $\psi$ is the incident photon flux. The ED absorption cross section integrated over the transition, $\sigma_{ij}^{ED}$, is calculated from $S_{ij}^{ED}$ using the relation:

$$\sigma_{int}^{ED} = \frac{8\pi^3 e^2 \tilde{\upsilon}}{3hcg_i}\frac{(n^2+2)^2}{9n}S_{ij}^{ED} \tag{S10}$$

Here, $\sigma_{int}^{ED}$ is the ED absorption cross section integrated over the entire transition. We handled two cases: resonant and non-resonant (phonon-assisted) absorption. For resonant absorption, we assumed that the laser excitation source has a width of 1 nm, and the absorption peak is a Gaussian curve centered at $\tilde{\upsilon}$. In this case, $\sigma_{ij}^{ED}$ is determined from $\sigma_{int}^{ED}$ using the expression:

$$\sigma_{ij} = \sigma_{int}^{ED} \frac{1}{w\sqrt{2\pi}} e^{-(v_{ex}-\tilde{v})^2/(2w^2)} \tag{S11}$$

where $v_{ex}$ is the energy (in wavenumbers) of the excitation source. $w$ is the linewidth of the absorption transition, related to the absorption FWHM by the expression: $w = FWHM_{abs}/(2\sqrt{2ln2})$. For $FWHM_{abs}$ see Supplementary Table S5.

For non-resonant absorption, the cross-section was modified to account for phonon assistance, as for MPR:

$$\sigma_{ij,PA} = \sigma_{ij,0}.e^{-\alpha|v_{ex}-\tilde{v}|} \tag{S12}$$

where $\sigma_{ij,0}$ is the value of the resonant $\sigma_{ij,0}$ when $v_{ex} = \tilde{v}$.

**Magnetic dipole (MD) radiation and absorption probabilities**. The MD radiative rate constant $A_{ij}^{MD}$ is calculated with the equations:

$$A_{ij}^{MD} = \frac{64\pi^4 e^2 \tilde{v}^3}{3h(2J+1)} n^3 S_{ij}^{MD} \tag{S13}$$

$$S_{ij}^{MD} = \mu_B^2 | < [SL]J_i||L+2S||[SL]J_j > |^2 \tag{S14}$$

where $\mu_B$ is the Bohr magneton. $L$ and $S$ are the angular momentum and spin operators, respectively. Note: these operators must be used on the eigenvectors of the intermediate coupled states of each lanthanide ion since each manifold total angular momentum $J$ is treated as a linear combination of various $LS$ states with the same $J$. The integrated MD absorption cross-section and oscillator strength are calculated using the expressions below:

$$\sigma_{\text{int}}^{MD} = \frac{\pi e^2}{mc^2} f_{ij}^{MD} \tag{S15}$$

$$f_{ij}^{MD} = \frac{8\pi^2 mc\tilde{v}}{3he^2 g_i} n S_{ij}^{MD} \tag{S16}$$

Here, $\sigma_{int}^{MD}$ is the MD absorption cross-section integrated over the entire transition. The actual MD absorption probability is calculated from $\sigma_{int}^{ED}$ in the same manner described for ED absorption above.

**Phonon-assisted energy transfer probabilities.** The orientation-averaged rate of dipole-dipole energy transfer $W_{ij,kl}^{ET}$ with donor transition $i \rightarrow j$ and acceptor transition $k \rightarrow l$ is described as a function of the inter-ion distance $R$ and the ED line strengths $S^{ED}$ of the donor and acceptor transitions[36]:

$$W_{ij,kl}^{ET} = \frac{C_{DA,ijkl}}{R^6} \quad \text{(S17)}$$

$$C_{DA,ijkl} = \frac{8\pi^2 e^4 s_0}{3h^2 c g_i g_k}(\frac{n^2+2}{3n})^4 S_{ij}^{ED} S_{kl}^{ED} \quad \text{(S18)}$$

In this expression, $C_{DA,ijkl}$ is the energy transfer micro-parameter, $g_i$ is the $2J+1$ number of states in manifold $i$, and $s_0$ is the overlap integral between the normalized donor emission spectrum and acceptor absorption spectrum. To account for phonon assistance during non-resonant energy transfer, the phonon-assisted energy transfer micro-parameter is used in place of the resonant ET micro-parameter and is calculated as[37]:

$$C_{DA}^{PAET} = C_{DA,ijkl} \cdot exp[-\beta_{MPR}\Delta E] \quad \text{(S19)}$$

where $\beta_{MPR} = \alpha - \ln(2)/\hbar\omega$, and $\Delta E$ is the net energy difference between the donor and acceptor transitions. $C_{DA,ijkl}$ is the energy transfer micro-parameter assuming a resonant ET process, calculated using eq. S18.

To account for energy transfer over all possible donor-acceptor distances *R* in a nanocrystal, the total phonon-assisted energy transfer (PAET) rate constant for a given donor can be calculated by integrating Eq. S17 over all *R* (ref: 38). In the case of a homogeneously doped material (see Supporting Information of ref: 32), we can define the PAET microparameter as:

$$W_{ij,kl,avg}^{PAET} = \frac{4\pi}{3} N_A C_{DA}^{PAET} R_{\min}^{-3} = N_A P_{ij,kl}^{MAET} \quad \text{(S20)}$$

To make the rate constant independent from the acceptor population, Eq. S6 uses the constant, $P_{ij,kl}^{MAET}$ (units of $cm^3$), whose relationship with $W_{ij,kl,avg}^{MAET}$ is shown above.

**Supplementary Table S4**. Non-radiative and radiative rate constants and energy transfer (ET) microparameters for the most prominent $Nd^{3+}$ transitions in $NaYF_4$ and $KPb_2Cl_5$ hosts. Note that the difference in MPR rate constants between the two host matrices has the greatest impact on the $Nd^{3+}$ photoactivation dynamics. [a] Transitions in parentheses denote complementary cross-relaxation (CR) transitions.

| Transition type | Transition $i \rightarrow j^a$ | $KPb_2Cl_5:Nd^{3+}$ | $NaYF_4:Nd^{3+}$ |
|---|---|---|---|
| | | Rate constant ($s^{-1}$) | |
| MPR | $^4I_{11/2} \rightarrow {^4I_{9/2}}$ | **185.85** | **654905.30** |
| | $^4I_{13/2} \rightarrow {^4I_{11/2}}$ | 98.39 | 397131.51 |
| | $^4I_{15/2} \rightarrow {^4I_{13/2}}$ | 4.92 | 236160.87 |
| | $^4F_{3/2} \rightarrow {^4I_{15/2}}$ | $2.78 \times 10^{-16}$ | 2.27 |
| ED abs. | $^4I_{9/2} \rightarrow {^4F_{3/2}}$ | 0.0023 | 0.227 |
| | $^4I_{11/2} \rightarrow {^4F_{3/2}}$ | 125.23 | 246.09 |
| ED em. | $^4F_{3/2} \rightarrow {^4I_J}$ (J = 9/2, 11/2, 13/2) | 4709.96 | 4502.91 |
| Cross-relaxation (CR) | | ET microparameter (x $10^{-40}$ $cm^6/s$) | |
| CR1 | $^4I_{13/2} \rightarrow {^4I_{11/2}}$ ($^4I_{9/2} \rightarrow {^4I_{11/2}}$) | 363 | 1598 |
| CR2 | $^4I_{15/2} \rightarrow {^4I_{13/2}}$ ($^4I_{9/2} \rightarrow {^4I_{11/2}}$) | 351 | 1217 |
| CR3 | $^4F_{3/2} \rightarrow {^4I_{15/2}}$ ($^4I_{9/2} \rightarrow {^4I_{15/2}}$) | 0.291 | n/a |
| CR4 | $^4F_{3/2} \rightarrow {^4I_{13/2}}$ ($^4I_{9/2} \rightarrow {^4I_{15/2}}$) | 0.564 | 1.75 |
| CR5 | $^4F_{3/2} \rightarrow {^4I_{15/2}}$ ($^4I_{9/2} \rightarrow {^4I_{13/2}}$) | 0.812 | 2.49 |

*Calculation methods*. We calculated the steady-state population of each $Nd^{3+}$ manifold under a given irradiation through numerical integration of the rate equations in Equation S6, which account for tens of thousands of simultaneous transitions. When performing power-dependent simulations, the population of $Nd^{3+}$ manifolds under a given power density value was simulated starting from the populations obtained at a prior power value. Cross-relaxation knock-out experiments were performed by ignoring transitions that met specified $i \rightarrow j$ & $l \rightarrow k$ criteria during the assembly of rate equations (Supplementary Figure S11)[31].

Kinetic calculations were performed according to the method of Chan et al. using Igor Pro 7 (Wavemetrics). N ordinary differential equations, which represent the population of each N manifold in the simulated system, were solved numerically using the Igor Pro Backwards Differentiation Formula integration method. Calculation parameters are given in Supplementary Tables S3 and S5.

**Supplementary Table S5.** General simulation parameters. [a] Simulation time period of 2000 ms ensures that the simulated system reaches a steady state for any excitation laser power density.

| Parameter | Value | Units |
|---|---|---|
| Simulation time period[a] | 2000 | ms |
| Radiative rate ($A_{ij}$) threshold | $10^{-4}$ | $s^{-1}$ |
| Phonon energy[1,6] | 250 | $cm^{-1}$ |
| $W^0_{MPR}$ rate constant[33] | $0.5 \cdot 10^7$ | $s^{-1}$ |
| $\alpha$ MPR rate constant[33] | $12 \cdot 10^{-3}$ | cm |
| Index of refraction | 1.96 | - |
| Volume per potential dopant site | 0.1082 | $nm^3$ |
| Minimum dopant distance | 0.4765 | nm |
| Absorption FWHM | 1000 | $cm^{-1}$ |
| Incident laser power density | $10^3$-$10^4$ | $W/cm^2$ |
| Incident excitation wavelength | 1064 | nm |

## 11. Mechanistic investigation of IOB

### 11.1 Additional knockout simulation data

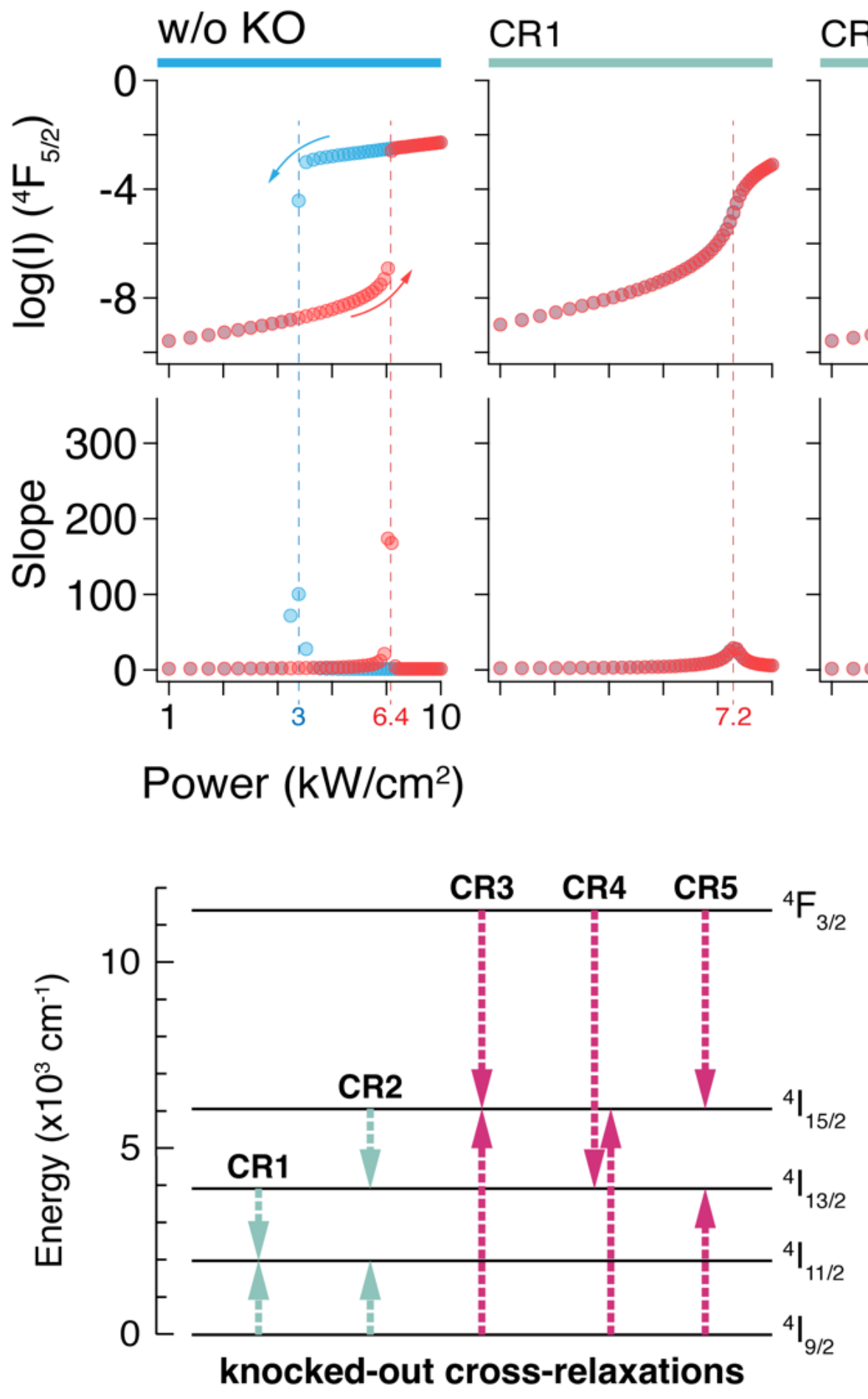

**Supplementary Figure S11.** Simulated pump (1064 nm) power dependence of $Nd^{3+}$ emission (810 nm, $^4F_{5/2} \rightarrow {}^4I_{9/2}$) in $KPb_2Cl_5$: 20 mol% $Nd^{3+}$ ANPs at 77 K. Red data points represent simulating with increasing powers (up scan) and blue with decreasing (down scan). Switch-on (red) and switch-off (blue) power density thresholds are also shown. Legend: w/o KO – wild-type simulation, knock-outs (KOs): CR1 – ($^4I_{9/2}\rightarrow{}^4I_{11/2}$):($^4I_{13/2}\rightarrow{}^4I_{11/2}$), CR2 – ($^4I_{9/2}\rightarrow{}^4I_{11/2}$):($^4I_{15/2}\rightarrow{}^4I_{13/2}$), CR3 – ($^4I_{9/2}\rightarrow{}^4I_{15/2}$):($^4F_{3/2}\rightarrow{}^4I_{15/2}$), CR4 – ($^4I_{9/2}\rightarrow{}^4I_{15/2}$):($^4F_{3/2}\rightarrow{}^4I_{13/2}$), CR5 – ($^4I_{9/2}\rightarrow{}^4I_{13/2}$):($^4F_{3/2}\rightarrow{}^4I_{15/2}$). The knocked-out cross-relaxation processes are shown in a simplified $Nd^{3+}$ energy level diagram below.

### 11.2 Supplementary Note S2. Results of a simplified rate equation model

To summarize the photon avalanching mechanism by the most critical transitions in $KPb_2Cl_5$: 20 mol% $Nd^{3+}$, we simulated the power dependence of the system using only the first 7 energy states of $Nd^{3+}$ ions (Supplementary Figure S12). The simulation accurately reproduced the experimental observations of luminescence hysteresis and population inversion and allowed to simplify the visualization of photo-activation mechanisms (Supplementary Figure S13).

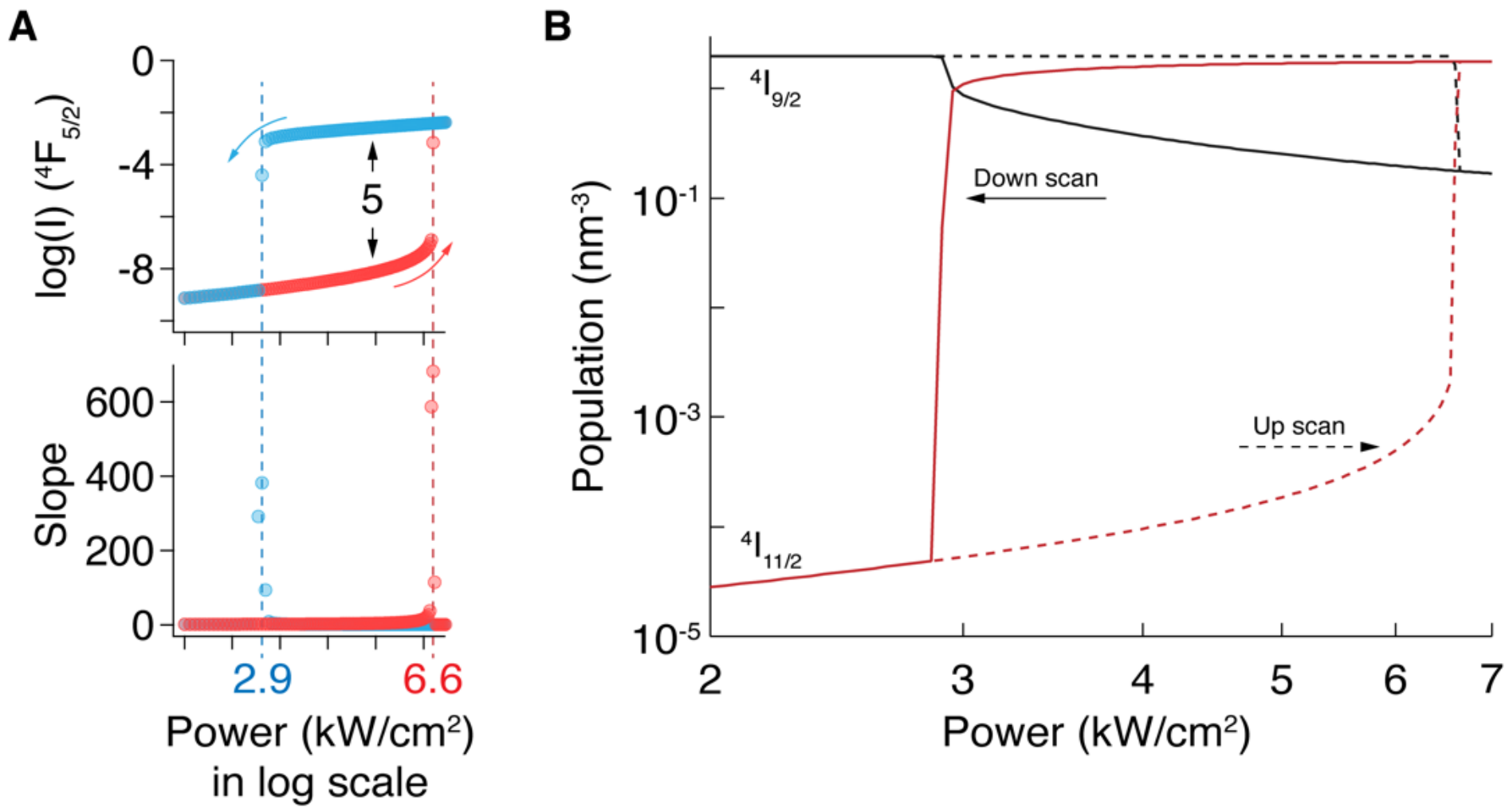

**Supplementary Figure S12. A** – Simulated pump (1064 nm) power dependence of $Nd^{3+}$ emission (810 nm, $^4F_{5/2} \rightarrow {}^4I_{9/2}$) in $KPb_2Cl_5$: 20 mol% $Nd^{3+}$ ANPs at 77 K, accounting only the first 7 energy levels (from $^4I_{9/2}$ to $^4F_{5/2}$) of $Nd^{3+}$ ions. Red data points represent simulating increasing powers (up scan) and blue with decreasing (down scan). The switch-on (red) and switch-off (blue) power density thresholds are also shown. Arrows indicate a power density (black) within the bistable region for which the photo-activation mechanisms are shown in Supplementary Figure S13. **B** – Simulated pump (1064 nm) power dependence of $Nd^{3+}$ ground ($^4I_{9/2}$) and first excited ($^4I_{11/2}$) states population in $KPb_2Cl_5$: 20 mol% $Nd^{3+}$ ANPs at 77 K, accounting only the first 7 energy levels (from $^4I_{9/2}$ to $^4F_{5/2}$) of $Nd^{3+}$ ions. Dashed lines indicate simulating with increasing powers (up scan) and solid with decreasing (down scan).

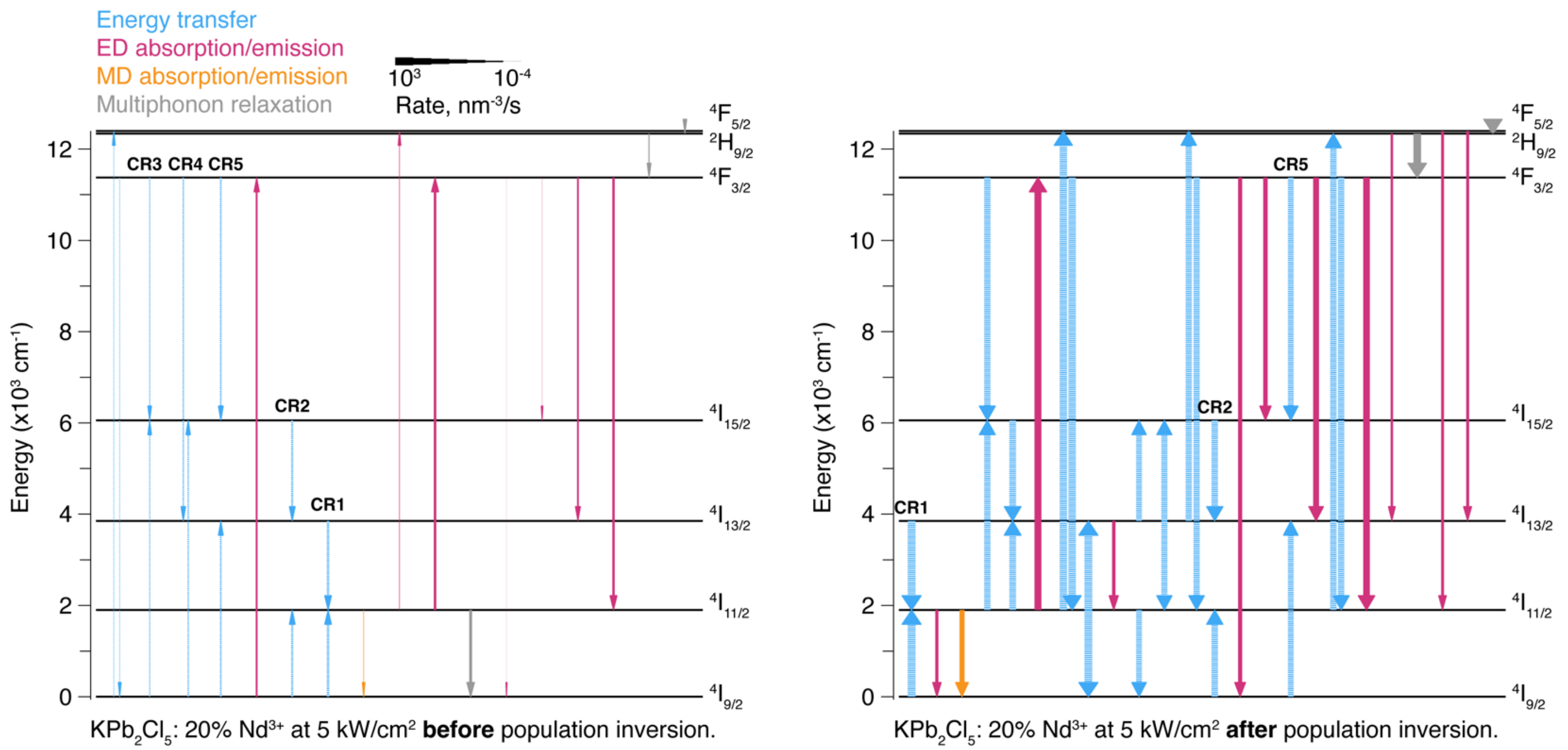

**Supplementary Figure S13.** Simplified energy level diagram and photophysical mechanism in $KPb_2Cl_5$: 20 mol% $Nd^{3+}$ ANPs at 77 K, accounting only for the first 7 energy levels (from $^4I_{9/2}$ to $^4F_{5/2}$) of $Nd^{3+}$ ions. The two diagrams represent primary energy transfer (blue arrows), photon absorption/emission (pink and yellow arrows), and multiphonon relaxation (gray arrows) processes in $KPb_2Cl_5$: 20 mol% $Nd^{3+}$ ANPs at the same 1064 nm pump power density of 5 kW/cm$^2$ when in the dark (left panel, before switch-on) or bright states (right panel, after switch-on).

### 11.3 Additional data for mechanistic investigations

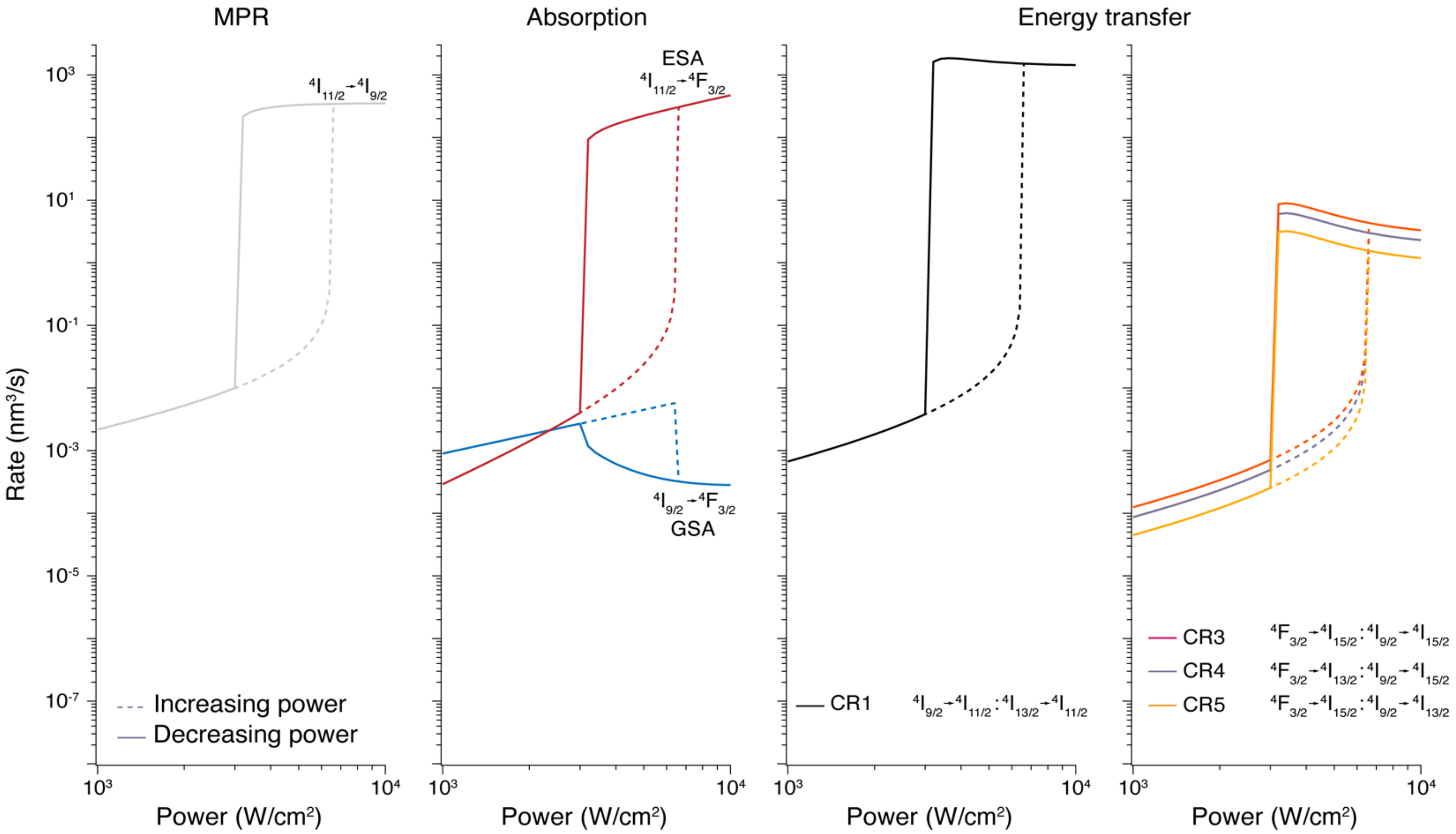

**Supplementary Figure S14.** Simulated pump (1064 nm) power dependence of different $Nd^{3+}$ transitions in $KPb_2Cl_5$: 20 mol% $Nd^{3+}$ ANPs. MPR – multiphoton relaxation, ESA – excited state absorption, GSA – ground state absorption, CR – cross-relaxation.

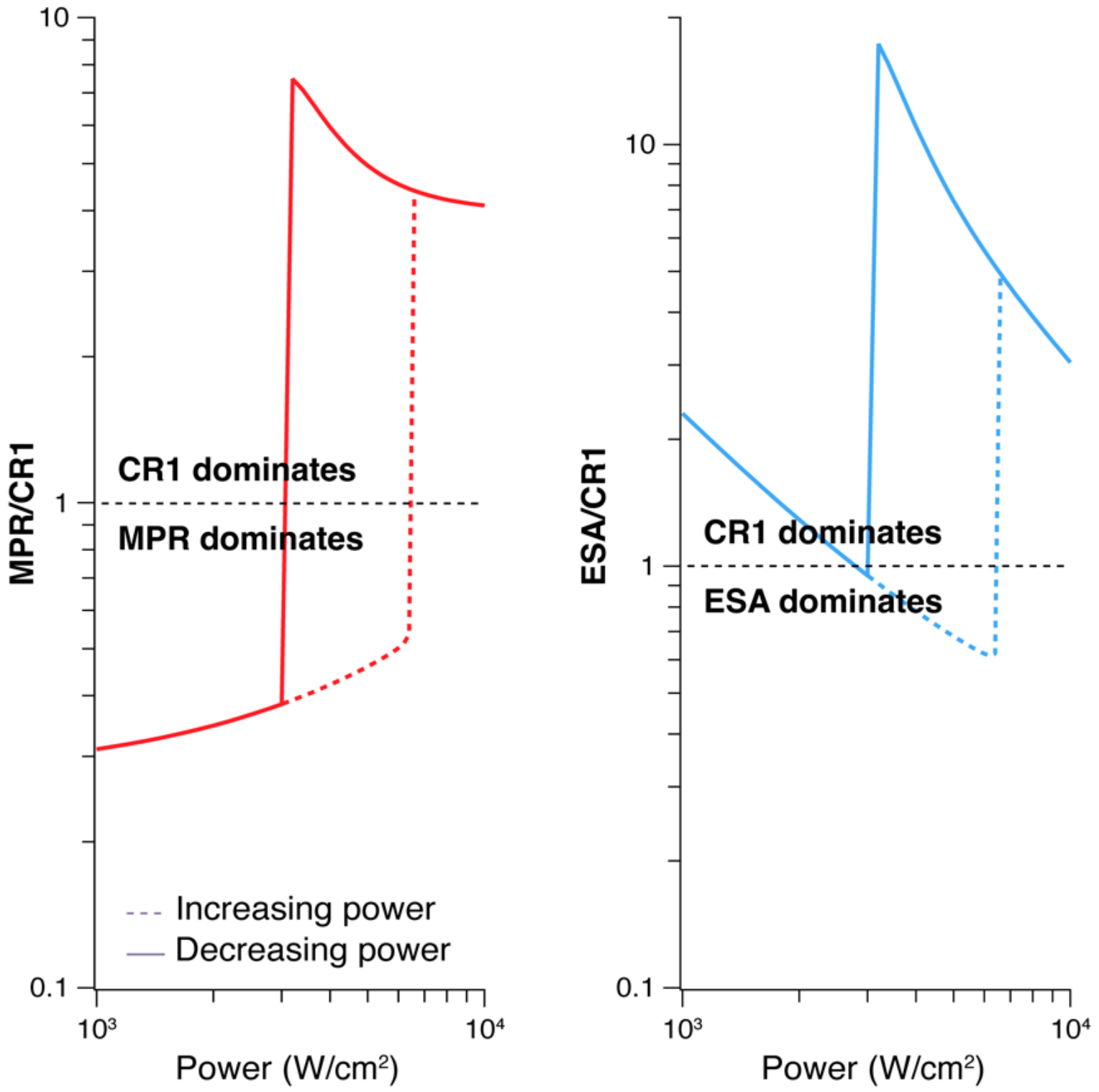

**Supplementary Figure S15.** Simulated pump (1064 nm) power-dependent ratio of MPR/CR1 (left) and ESA/CR1 (right) in $KPb_2Cl_5$: 20 mol% $Nd^{3+}$ ANPs. When the bistable $KPb_2Cl_5$:$Nd^{3+}$ ANPs are in their bright state, the CR1 dominates over MPR (left panel). Similarly, to sustain a positive feedback loop that leads to bistability, ESA and CR1 tradeoff in their contribution. Before the switch-on threshold, the probability of ESA increases with increasing pump power density and is the determining factor that switches the bistable ANPs from their dark to a bright state (right panel, dashed line). However, in the bright state, CR1 dominates over ESA, and with decreasing pump powers, compensates for the diminishing rate of ESA.

*Pathway analysis*. To determine the most critical transitions for a given energy transfer pathway, we considered only transitions with the highest branching fractions ($\beta_{ij}$) or contribution fractions ($\kappa_{ij}$), where:

$$\beta_{ij,t} = \frac{\left(\frac{dN_i}{dt}\right)_{i\to j}^{t}}{\sum_{t=ED,MD,ET,MPR}\sum_m \left(\frac{dN_i}{dt}\right)_{i\to m}^{t}} \tag{S21}$$

$$\kappa_{ij,t} = \frac{\left(\frac{dN_j}{dt}\right)_{i\to j}^{t}}{\sum_t\sum_m \left(\frac{dN_j}{dt}\right)_{m\to j}^{t}} \tag{S22}$$

For a given transition type $t$ (ED/MD radiative, ET, MPR) from $i$ to $j$, $\kappa_{ij}$ ($\beta_{ij}$) represents that transition's fractional contribution to the overall population (depletion) of $j$ ($i$) by all types of transitions to (from) level $j$ ($i$). The rates $dN_i/dt$ are determined by the populations of the originating states and the corresponding microscopic rate constants, as shown by the individual components of Equation S6. To select the most significant incoming transitions for each manifold $i$, we retained the transitions with the highest $\kappa_{ij}$ that cumulatively contributed to over 90% of the respective population of $j$ (i.e., $\sum \kappa_{ij}$ > 0.90). The most significant depopulating transitions can be filtered similarly using $\beta_{ij}$ values.

The most important contributing and branching transitions in $KPb_2Cl_5$: 20 mol% $Nd^{3+}$ before (dark state) and after (bright state) the switch-on power threshold are summarized in Supplementary Tables S6 and S7.

**Supplementary Table S6**. Major contributing transitions for $KPb_2Cl_5$: 20 mol% $Nd^{3+}$ calculated at 5 kW/cm$^2$ at a steady state within the bistable region of luminescence before the switch-on threshold (power up) and after (power down). [a] Transitions in parentheses denote complementary ET transitions. [b] n/s – not significant.

| | | | Power up | | Power down | |
|---|---|---|---|---|---|---|
| Level $i$ | Transition type | Transition $i \rightarrow j$[a] | Rate (nm$^3$/s) | $\kappa_{ij}$ | Rate (nm$^3$/s) | $\kappa_{ij}$ |
| $^4I_{9/2}$ | MPR | $^4I_{11/2} \rightarrow {}^4I_{9/2}$ | 0.0357 | 0.9626 | 330.7373 | 0.1829 |
| | ET acceptor | $^4I_{11/2} \rightarrow {}^4I_{9/2}$ ($^4I_{11/2} \rightarrow {}^4I_{13/2}$) | n/s[b] | n/s | 1408.058 | 0.7788 |
| $^4I_{11/2}$ | ET acceptor | $^4I_{9/2} \rightarrow {}^4I_{11/2}$ ($^4I_{13/2} \rightarrow {}^4I_{11/2}$) | 0.0164 | 0.2688 | 1638.3481 | 0.4181 |
| | ET donor | $^4I_{13/2} \rightarrow {}^4I_{11/2}$ ($^4I_{9/2} \rightarrow {}^4I_{11/2}$) | 0.0164 | 0.2688 | 1638.3481 | 0.4181 |
| | ET donor | $^4F_{3/2} \rightarrow {}^4I_{11/2}$ ($^4I_{11/2} \rightarrow {}^4F_{5/2}$) | n/s | n/s | 170.1308 | 0.0434 |
| | ET acceptor | $^4I_{9/2} \rightarrow {}^4I_{11/2}$ ($^4I_{15/2} \rightarrow {}^4I_{13/2}$) | 0.0068 | 0.1118 | n/s | n/s |
| | ED em. | $^4F_{3/2} \rightarrow {}^4I_{11/2}$ | 0.0190 | 0.3109 | 43.0233 | 0.0110 |
| $^4I_{13/2}$ | ET acceptor | $^4I_{9/2} \rightarrow {}^4I_{13/2}$ ($^4F_{3/2} \rightarrow {}^4I_{15/2}$) | 0.0030 | 0.1841 | n/s | n/s |
| | ET donor | $^4I_{15/2} \rightarrow {}^4I_{13/2}$ ($^4I_{9/2} \rightarrow {}^4I_{11/2}$) | 0.0068 | 0.4152 | 56.9246 | 0.0300 |
| | ET acceptor | $^4I_{11/2} \rightarrow {}^4I_{13/2}$ ($^4I_{15/2} \rightarrow {}^4I_{13/2}$) | n/s | n/s | 106.4997 | 0.0562 |
| | ET donor | $^4I_{15/2} \rightarrow {}^4I_{13/2}$ ($^4I_{11/2} \rightarrow {}^4I_{13/2}$) | n/s | n/s | 106.4997 | 0.0562 |
| | ET acceptor | $^4I_{11/2} \rightarrow {}^4I_{13/2}$ ($^4I_{11/2} \rightarrow {}^4I_{9/2}$) | n/s | n/s | 1408.0580 | 0.7424 |
| | ED em. | $^4F_{3/2} \rightarrow {}^4I_{13/2}$ | 0.0036 | 0.2161 | 8.0558 | 0.0107 |
| $^4I_{15/2}$ | ET donor | $^4F_{3/2} \rightarrow {}^4I_{15/2}$ ($^4I_{9/2} \rightarrow {}^4I_{15/2}$) | 0.0011 | 0.1447 | n/s | n/s |
| | ET acceptor | $^4I_{9/2} \rightarrow {}^4I_{15/2}$ ($^4F_{3/2} \rightarrow {}^4I_{13/2}$) | 0.0021 | 0.2802 | n/s | n/s |
| | ET donor | $^4F_{3/2} \rightarrow {}^4I_{15/2}$ ($^4I_{9/2} \rightarrow {}^4I_{13/2}$) | 0.0030 | 0.4034 | 5.8307 | 0.0337 |
| | ET donor | $^4F_{3/2} \rightarrow {}^4I_{15/2}$ ($^4F_{3/2} \rightarrow {}^4G_{5/2}$) | n/s | n/s | 13.5187 | 0.0781 |
| | ET acceptor | $^4I_{13/2} \rightarrow {}^4I_{15/2}$ ($^4I_{11/2} \rightarrow {}^4I_{9/2}$) | n/s | n/s | 21.4445 | 0.1239 |
| | ET acceptor | $^4I_{11/2} \rightarrow {}^4I_{15/2}$ ($^4F_{3/2} \rightarrow {}^4I_{15/2}$) | n/s | n/s | 23.8860 | 0.1380 |
| | ET donor | $^4F_{3/2} \rightarrow {}^4I_{15/2}$ ($^4I_{11/2} \rightarrow {}^4I_{15/2}$) | n/s | n/s | 23.8860 | 0.1380 |
| | ET acceptor | $^4I_{13/2} \rightarrow {}^4I_{15/2}$ ($^4I_{13/2} \rightarrow {}^4I_{11/2}$) | n/s | n/s | 46.6821 | 0.2698 |
| | ED em. | $^4F_{3/2} \rightarrow {}^4I_{15/2}$ | 0.0002 | 0.0231 | 0.3935 | 0.0023 |
| $^4F_{3/2}$ | MPR | $^2H_{9/2} \rightarrow {}^4F_{3/2}$ | 0.0015 | 0.0464 | 511.3653 | 0.6775 |
| | ED abs. | $^4I_{9/2} \rightarrow {}^4F_{3/2}$ | 0.0045 | 0.1428 | n/s | n/s |
| | ED abs. | $^4I_{11/2} \rightarrow {}^4F_{3/2}$ | 0.0241 | 0.7653 | 222.8575 | 0.2952 |
| $^2H_{9/2}$ | MPR | $^4F_{5/2} \rightarrow {}^2H_{9/2}$ | 0.0012 | 0.8478 | 462.8812 | 0.9008 |
| $^4F_{5/2}$ | MPR | $^4F_{7/2} \rightarrow {}^4F_{5/2}$ | 0.0002 | 0.1623 | 220.2116 | 0.4722 |
| | ET acceptor | $^4I_{9/2} \rightarrow {}^4F_{5/2}$ ($^4F_{3/2} \rightarrow {}^4I_{9/2}$) | 0.0005 | 0.4296 | n/s | n/s |
| | ET acceptor | $^4F_{3/2} \rightarrow {}^4F_{5/2}$ ($^4I_{13/2} \rightarrow {}^4I_{11/2}$) | n/s | n/s | 17.5881 | 0.0377 |
| | ET acceptor | $^4I_{11/2} \rightarrow {}^4F_{5/2}$ ($^4F_{3/2} \rightarrow {}^4I_{11/2}$) | n/s | n/s | 170.1308 | 0.3648 |
| | ED abs. | $^4I_{11/2} \rightarrow {}^4F_{5/2}$ | 0.0004 | 0.3553 | n/s | n/s |
| $^4F_{7/2}$ | MPR | $^4S_{3/2} \rightarrow {}^4F_{7/2}$ | 0.0002 | 0.7434 | 97.9593 | 0.4408 |
| | ET donor | $^2H_{11/2} \rightarrow {}^4F_{7/2}$ ($^4I_{9/2} \rightarrow {}^4I_{11/2}$) | 0.00002 | 0.1052 | n/s | n/s |
| | ET donor | $^4F_{9/2} \rightarrow {}^4F_{7/2}$ ($^4I_{11/2} \rightarrow {}^4I_{13/2}$) | n/s | n/s | 2.6266 | 0.0118 |
| | ET acceptor | $^4F_{3/2} \rightarrow {}^4F_{7/2}$ ($^4I_{11/2} \rightarrow {}^4I_{9/2}$) | n/s | n/s | 10.7590 | 0.0484 |
| | ET acceptor | $^4F_{3/2} \rightarrow {}^4F_{7/2}$ ($^4I_{13/2} \rightarrow {}^4I_{11/2}$) | n/s | n/s | 22.3553 | 0.1006 |
| | ET acceptor | $^4I_{13/2} \rightarrow {}^4F_{7/2}$ ($^4F_{3/2} \rightarrow {}^4I_{11/2}$) | n/s | n/s | 31.8461 | 0.1433 |
| $^4S_{3/2}$ | MPR | $^4F_{9/2} \rightarrow {}^4S_{3/2}$ | 0.0001 | 0.8966 | 59.1993 | 0.6017 |
| | ET acceptor | $^4I_{13/2} \rightarrow {}^4S_{3/2}$ ($^4F_{3/2} \rightarrow {}^4I_{11/2}$) | n/s | n/s | 23.6961 | 0.2408 |

**Supplementary Table S7**. Major branching transitions for $KPb_2Cl_5$: 20 mol% $Nd^{3+}$ calculated at 5 $kW/cm^2$ at a steady state within the bistable region of luminescence before the switch-on threshold (power up) and after (power down). [a] Transitions in parentheses denote complementary ET transitions. [b] n/s – not significant.

| | | | Power up | | Power down | |
|---|---|---|---|---|---|---|
| Level *i* | Transition type | Transition $i \rightarrow j$ [a] | Rate ($nm^3/s$) | $\beta_{ij}$ | Rate ($nm^3/s$) | $\beta_{ij}$ |
| $^4I_{9/2}$ | ET acceptor | $^4I_{9/2} \rightarrow {}^4F_{5/2}$ ($^4F_{3/2} \rightarrow {}^4I_{9/2}$) | 0.0005 | 0.0144 | n/s[b] | n/s |
| | ET acceptor | $^4I_{9/2} \rightarrow {}^4I_{15/2}$ ($^4F_{3/2} \rightarrow {}^4I_{13/2}$) | 0.0021 | 0.0567 | n/s | n/s |
| | ET acceptor | $^4I_{9/2} \rightarrow {}^4I_{13/2}$ ($^4F_{3/2} \rightarrow {}^4I_{15/2}$) | 0.0030 | 0.0816 | n/s | n/s |
| | ET acceptor | $^4I_{9/2} \rightarrow {}^4I_{11/2}$ ($^4I_{15/2} \rightarrow {}^4I_{13/2}$) | 0.0068 | 0.1840 | n/s | n/s |
| | ET acceptor | $^4I_{9/2} \rightarrow {}^4I_{11/2}$ ($^4I_{13/2} \rightarrow {}^4I_{11/2}$) | 0.0164 | 0.4423 | 1638.3481 | 0.9062 |
| | ED abs. | $^4I_{9/2} \rightarrow {}^4F_{3/2}$ | 0.0045 | 0.1211 | n/s | n/s |
| $^4I_{11/2}$ | MPR | $^4I_{11/2} \rightarrow {}^4I_{9/2}$ | 0.0357 | 0.5852 | 330.7373 | 0.0844 |
| | ET acceptor | $^4I_{11/2} \rightarrow {}^4F_{5/2}$ ($^4F_{3/2} \rightarrow {}^4I_{11/2}$) | n/s | n/s | 170.1308 | 0.0434 |
| | ET acceptor | $^4I_{11/2} \rightarrow {}^4I_{13/2}$ ($^4I_{11/2} \rightarrow {}^4I_{9/2}$) | n/s | n/s | 1408.0580 | 0.3593 |
| | ET donor | $^4I_{11/2} \rightarrow {}^4I_{9/2}$ ($^4I_{11/2} \rightarrow {}^4I_{13/2}$) | n/s | n/s | 1408.0580 | 0.3593 |
| | ED abs. | $^4I_{11/2} \rightarrow {}^4F_{3/2}$ | 0.0241 | 0.3943 | 222.8575 | 0.0569 |
| $^4I_{13/2}$ | ET acceptor | $^4I_{13/2} \rightarrow {}^4I_{15/2}$ ($^4I_{13/2} \rightarrow {}^4I_{11/2}$) | n/s | n/s | 46.6821 | 0.0246 |
| | ET donor | $^4I_{13/2} \rightarrow {}^4I_{11/2}$ ($^4I_{9/2} \rightarrow {}^4I_{11/2}$) | 0.0164 | 0.9979 | 1638.3481 | 0.8638 |
| | ET acceptor | $^4I_{13/2} \rightarrow {}^4F_{7/2}$ ($^4F_{3/2} \rightarrow {}^4I_{11/2}$) | n/s | n/s | 31.8461 | 0.0168 |
| $^4I_{15/2}$ | ET donor | $^4I_{15/2} \rightarrow {}^4I_{13/2}$ ($^4I_{9/2} \rightarrow {}^4I_{11/2}$) | 0.0068 | 0.9095 | 56.9246 | 0.3289 |
| | ET donor | $^4I_{15/2} \rightarrow {}^4I_{13/2}$ ($^4I_{11/2} \rightarrow {}^4I_{13/2}$) | n/s | n/s | 106.4997 | 0.6154 |
| $^4F_{3/2}$ | ET acceptor | $^4F_{3/2} \rightarrow {}^4G_{5/2}$ ($^4F_{3/2} \rightarrow {}^4I_{15/2}$) | n/s | n/s | 13.5187 | 0.0179 |
| | ET donor | $^4F_{3/2} \rightarrow {}^4I_{15/2}$ ($^4F_{3/2} \rightarrow {}^4G_{5/2}$) | n/s | n/s | 13.5187 | 0.0179 |
| | ET donor | $^4F_{3/2} \rightarrow {}^4I_{13/2}$ ($^4F_{3/2} \rightarrow {}^4G_{7/2}$) | n/s | n/s | 39.3544 | 0.0521 |
| | ET acceptor | $^4F_{3/2} \rightarrow {}^4G_{7/2}$ ($^4F_{3/2} \rightarrow {}^4I_{13/2}$) | n/s | n/s | 39.3544 | 0.0521 |
| | ET acceptor | $^4F_{3/2} \rightarrow {}^4F_{7/2}$ ($^4I_{11/2} \rightarrow {}^4I_{9/2}$) | n/s | n/s | 10.7590 | 0.0143 |
| | ET donor | $^4F_{3/2} \rightarrow {}^4I_{11/2}$ ($^4I_{13/2} \rightarrow {}^4S_{3/2}$) | n/s | n/s | 23.6961 | 0.0314 |
| | ET donor | $^4F_{3/2} \rightarrow {}^4I_{11/2}$ ($^4I_{13/2} \rightarrow {}^4F_{7/2}$) | n/s | n/s | 31.8461 | 0.0422 |
| | ET donor | $^4F_{3/2} \rightarrow {}^4I_{11/2}$ ($^4I_{11/2} \rightarrow {}^4F_{5/2}$) | n/s | n/s | 170.1308 | 0.2254 |
| | ET donor | $^4F_{3/2} \rightarrow {}^4I_{15/2}$ ($^4I_{9/2} \rightarrow {}^4I_{15/2}$) | 0.0011 | 0.0345 | | |
| | ET donor | $^4F_{3/2} \rightarrow {}^4I_{15/2}$ ($^4I_{9/2} \rightarrow {}^4I_{13/2}$) | 0.0030 | 0.0963 | 5.8307 | 0.0077 |
| | ET acceptor | $^4F_{3/2} \rightarrow {}^4G_{9/2}$ ($^4F_{3/2} \rightarrow {}^4I_{13/2}$) | n/s | n/s | 35.1088 | 0.0465 |
| | ET donor | $^4F_{3/2} \rightarrow {}^4I_{13/2}$ ($^4F_{3/2} \rightarrow {}^4G_{9/2}$) | n/s | n/s | 35.1088 | 0.0465 |
| | ET acceptor | $^4F_{3/2} \rightarrow {}^4F_{5/2}$ ($^4I_{13/2} \rightarrow {}^4I_{11/2}$) | n/s | n/s | 17.5881 | 0.0233 |
| | ET donor | $^4F_{3/2} \rightarrow {}^4I_{11/2}$ ($^4I_{13/2} \rightarrow {}^4F_{5/2}$) | n/s | n/s | 20.5211 | 0.0272 |
| | ET acceptor | $^4F_{3/2} \rightarrow {}^4F_{7/2}$ ($^4I_{13/2} \rightarrow {}^4I_{11/2}$) | n/s | n/s | 22.3553 | 0.0296 |
| | ET donor | $^4F_{3/2} \rightarrow {}^4I_{15/2}$ ($^4I_{11/2} \rightarrow {}^4I_{15/2}$) | n/s | n/s | 23.8860 | 0.0316 |
| | ET donor | $^4F_{3/2} \rightarrow {}^4I_{11/2}$ ($^4I_{11/2} \rightarrow {}^2H_{9/2}$) | n/s | n/s | 23.9986 | 0.0318 |
| | ED em. | $^4F_{3/2} \rightarrow {}^4I_{9/2}$ | 0.0001 | 0.0038 | 0.2725 | 0.0004 |
| | ED em. | $^4F_{3/2} \rightarrow {}^4I_{11/2}$ | 0.0190 | 0.6034 | 43.0233 | 0.0570 |
| | ED em. | $^4F_{3/2} \rightarrow {}^4I_{13/2}$ | 0.0036 | 0.1130 | 8.0558 | 0.0107 |
| | ED em. | $^4F_{3/2} \rightarrow {}^4I_{15/2}$ | 0.0002 | 0.0055 | 0.3935 | 0.0005 |
| $^2H_{9/2}$ | MPR | $^2H_{9/2} \rightarrow {}^4F_{3/2}$ | 0.0015 | 0.9991 | 511.3653 | 0.9952 |
| $^4F_{5/2}$ | MPR | $^4F_{5/2} \rightarrow {}^2H_{9/2}$ | 0.0012 | 0.9977 | 462.8812 | 0.9926 |
| $^4F_{7/2}$ | MPR | $^4F_{7/2} \rightarrow {}^4F_{5/2}$ | 0.0002 | 0.9969 | 220.2116 | 0.9909 |
| $^4S_{3/2}$ | MPR | $^4S_{3/2} \rightarrow {}^4F_{7/2}$ | 0.0002 | 0.9995 | 97.9593 | 0.9956 |
| $^4F_{7/2}$ | MPR | $^4F_{9/2} \rightarrow {}^4S_{3/2}$ | 0.0001 | 0.8404 | 59.1993 | 0.6776 |
| | ET donor | $^4F_{9/2} \rightarrow {}^4F_{5/2}$ ($^4I_{11/2} \rightarrow {}^4I_{13/2}$) | n/s | n/s | 3.5064 | 0.0401 |
| | ET donor | $^4F_{9/2} \rightarrow {}^4I_{13/2}$ ($^4I_{11/2} \rightarrow {}^4F_{7/2}$) | n/s | n/s | 2.3896 | 0.0274 |
| | ET donor | $^4F_{9/2} \rightarrow {}^4I_{13/2}$ ($^4I_{11/2} \rightarrow {}^4F_{5/2}$) | n/s | n/s | 2.5531 | 0.0292 |
| | ET donor | $^4F_{9/2} \rightarrow {}^4F_{7/2}$ ($^4I_{11/2} \rightarrow {}^4I_{13/2}$) | n/s | n/s | 2.6266 | 0.0301 |

## 12. Temporal modulation of IOB

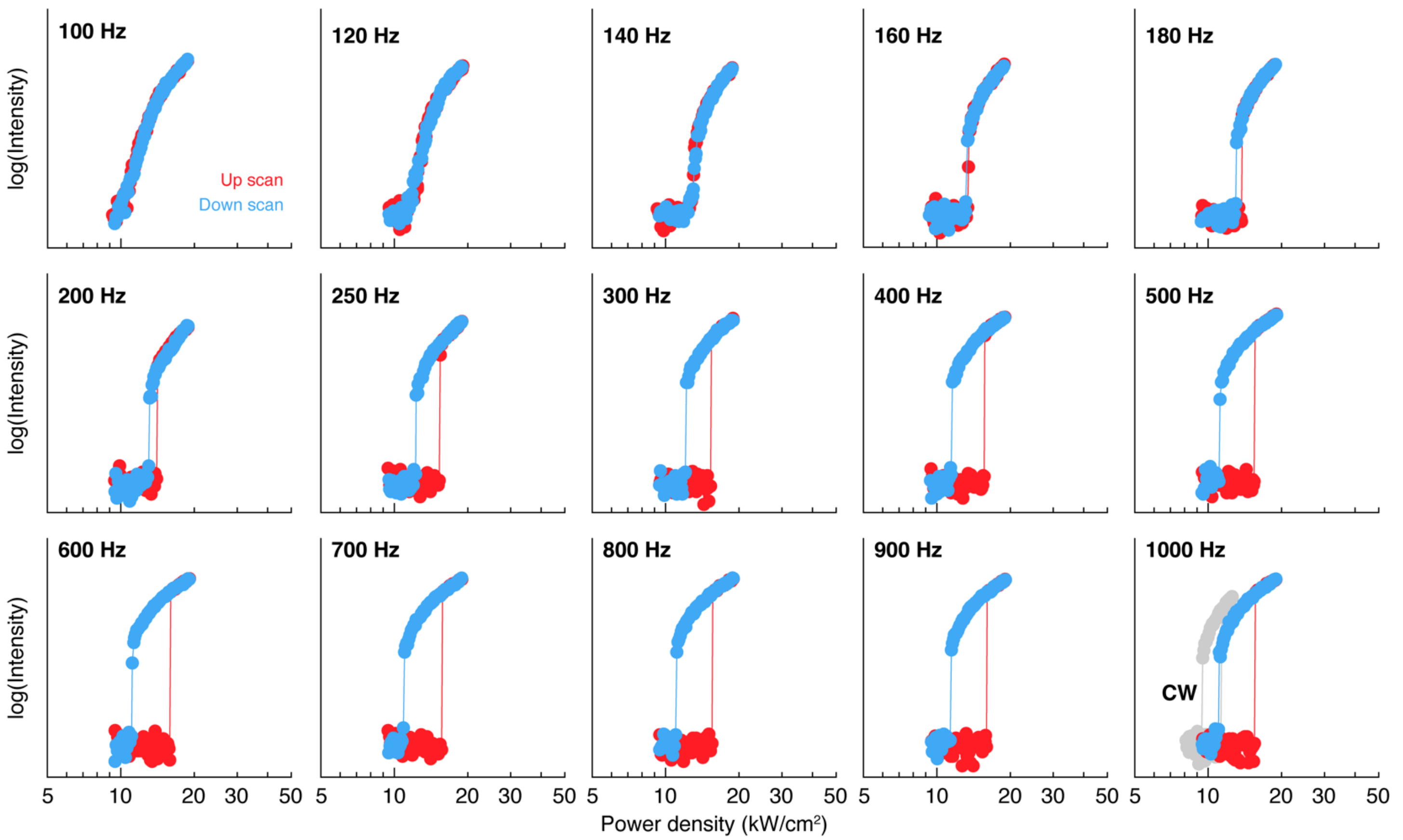

**Supplementary Figure S16.** Representative average pump (1064 nm) power dependence curves of $Nd^{3+}$ emission (810 nm, $^4F_{5/2} \rightarrow {}^4I_{9/2}$) in $KPb_2Cl_5:Nd^{3+}$ ANPs at 77 K under pulsed pump of different frequencies (40% duty cycle). CW – continuous wave data are shown for comparison in a 1000 Hz panel. Red data points represent scanning with increasing powers (up scan) and blue with decreasing (down scan).

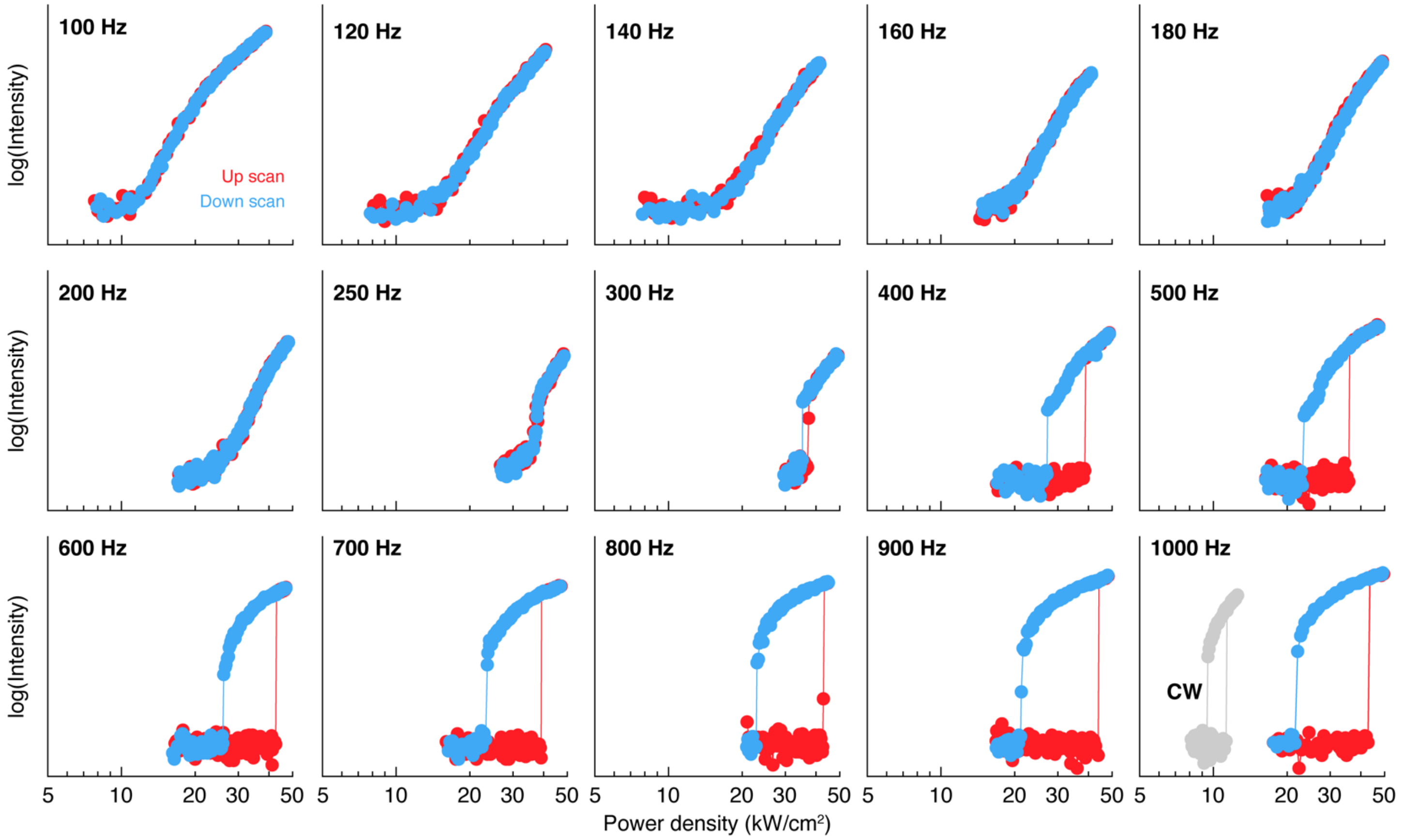

**Supplementary Figure S17.** Representative average pump (1064 nm) power dependence curves of $Nd^{3+}$ emission (810 nm, $^4F_{5/2} \rightarrow {}^4I_{9/2}$) in $KPb_2Cl_5:Nd^{3+}$ ANPs at 77 K under pulsed pump of different frequencies (10% duty cycle). CW – continuous wave data are shown for comparison in a 1000 Hz panel. Red data points represent scanning with increasing powers (up scan) and blue with decreasing (down scan).

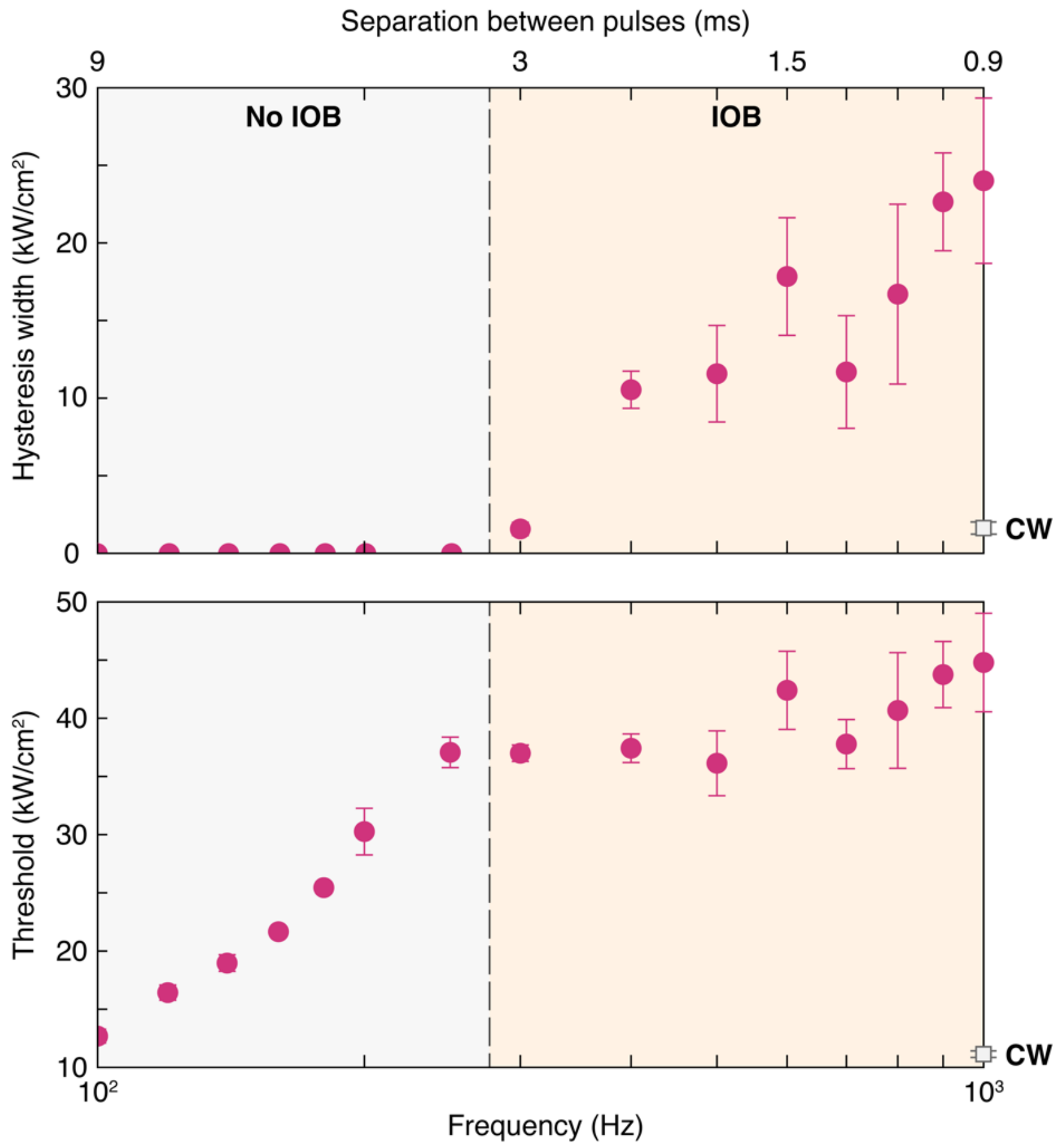

**Supplementary Figure S18.** Hysteresis width and photon avalanche switch-on threshold vs pump pulse frequency (10% duty cycle). Data is summarized from the Supplementary Figure S17. The corresponding time separation between pulses is also shown. Respective values under the CW pump are shown as squares. The data points are shown as the mean values ± 1 standard deviation ($n$ = 3).

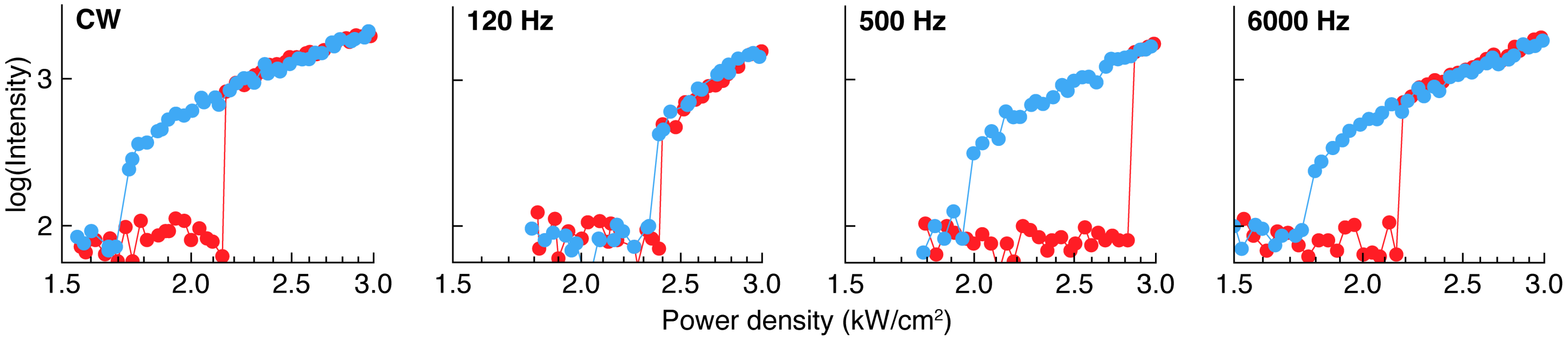

**Supplementary Figure S19.** Representative average pump (1064 nm) power dependence curves of $Nd^{3+}$ emission (810 nm, $^4F_{5/2} \rightarrow {}^4I_{9/2}$) in $KPb_2Cl_5:Nd^{3+}$ ANPs at 77 K under continuous wave and the pulsed pump of different frequencies (50% duty cycle). Red data points represent scanning with increasing powers (up scan) and blue with decreasing (down scan).

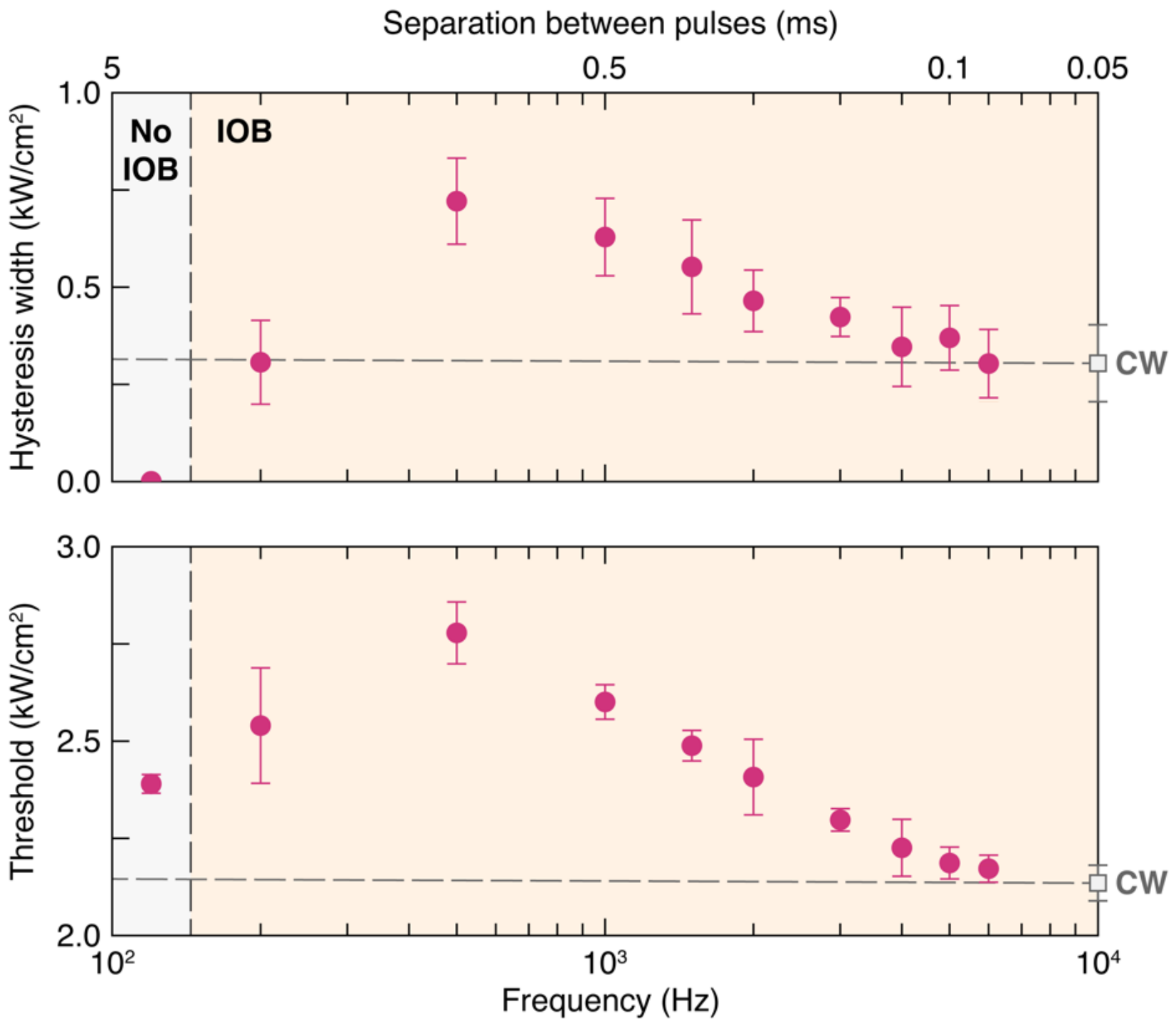

**Supplementary Figure S20.** Hysteresis width and photon avalanche switch ON threshold vs pump pulse frequency (50% duty cycle). Data is summarized from Figure S19. The corresponding time separation between pulses is also shown. At higher pump pulse frequencies, >4000 Hz, a quasi-CW regime is reached, and luminescence hysteresis of the $KPb_2Cl_5:Nd^{3+}$ ANPs matches that measured under the CW pump (shown as squares). The data points are shown as the mean values ± 1 standard deviation (*n* = 3).

## Supplementary Note S3. Time-resolved measurements of ANPs

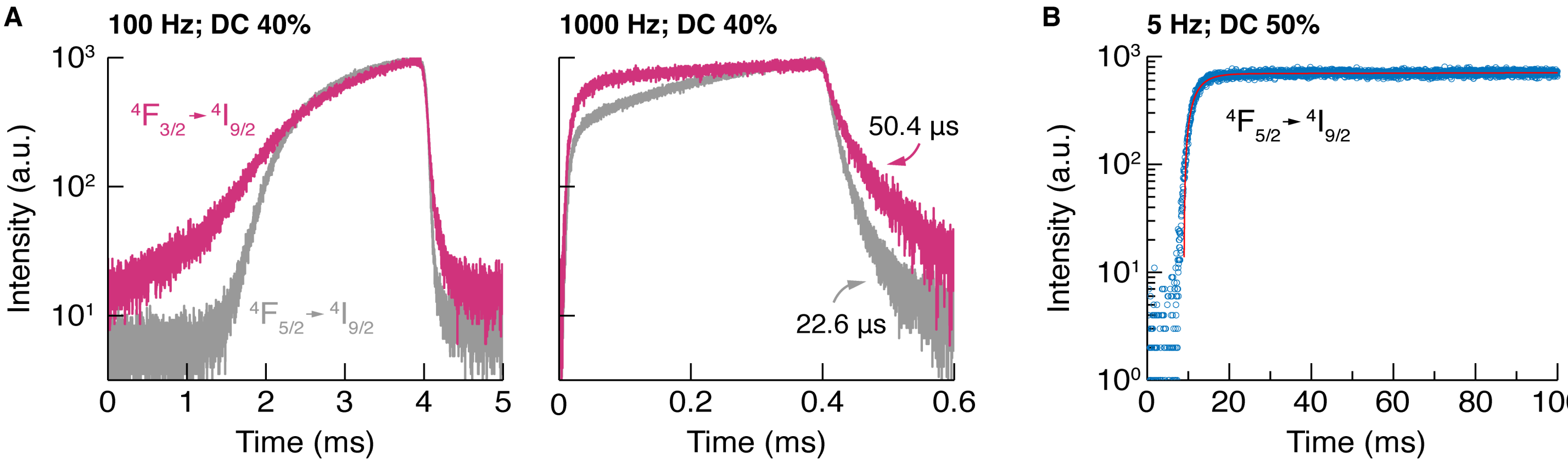

**Supplementary Figure S21.** Time-resolved photoluminescence of $KPb_2Cl_5:Nd^{3+}$ ANPs at 77 K measured for 880 nm ($^4F_{3/2} \rightarrow {}^4I_{9/2}$) and 810 nm ($^4F_{5/2} \rightarrow {}^4I_{9/2}$) optical transitions under the pulsed 1064 nm pump of different frequencies. The pulsed pump duty cycle was constant at 40%. The pulse frequencies of 100 Hz and 1000 Hz correspond to $KPb_2Cl_5:Nd^{3+}$ nanocrystals responding as regular or bistable ANPs, respectively (see Supplementary Figure S16 for power-dependent data). In the bistable excitation regime (1000 Hz), the rise time for populating $^4F_{3/2}$ energy level is 305 μs (calculated at 95% intensity from the steady state), and the decay times are 50.4 μs and 22.6 μs for $^4F_{3/2}$ and $^4F_{5/2}$ energy levels, respectively. The decay times were calculated by integrating the area under decay curves[31]. The estimated decay times are on the order of those measured for highly doped $NaGdF_4:Nd^{3+}$ and $LaPO_4:Nd^{3+}$ nanocrystals[41,42]. **B** - Time-resolved photoluminescence of $KPb_2Cl_5:Nd^{3+}$ ANPs at 77 K measured for 810 nm ($^4F_{5/2} \rightarrow {}^4I_{9/2}$) optical transition under a slow-pulsed 1064 nm pump (pulse frequency 5Hz, duty cycle 50%). Note, the measurement spot on the film sample is different than in **A**. The rise time for populating $^4F_{5/2}$ energy level is 20.8 ms (calculated at 95% intensity from the steady state). Blue circles are the experimental data points, and the red line is a double-exponential fit. Time-resolved photoluminescence was measured above the switch-on threshold at an average pump power intensity of 18.7 kW/cm$^2$ in **A** and 2.0 kW/cm$^2$ in **B**.

To better understand how pulsed modulation alters the luminescence hysteresis of $KPb_2Cl_5:Nd^{3+}$ ANPs, we performed time-resolved measurements for different pulsed excitation conditions (Supplementary Figure S21A). We note that with high-frequency pulsed excitation (e.g., 1000 Hz), emission of optically bistable ANPs is detected only in their bright state after the excitation threshold is reached (Supplementary Figure S16, 1000 Hz).

At high-frequency pulsed excitation, the population inversion (PI) of $^4I_{9/2}$ and $^4I_{11/2}$ energy levels persists between pulses, acting like a memory reservoir[43]. The time-resolved data (pink traces in Supplementary Figure S21) thus reflects the population of the $^4F_{3/2}$ emitting energy level directly via ESA, omitting the initial step of GSA. The observed delay in populating the $^4F_{5/2}$ energy level (gray traces in Supplementary Figure S21) results from the additional energy-transfer upconversion (Supplementary Figure S13).

In contrast, with low-frequency excitation pulses (e.g., 100 Hz) spaced further apart than the duration of PI, we observe a population of excited energy levels through a combination of GSA and ESA with each pulse. As a result, it takes several milliseconds for the ANPs to reach a quasi-steady state. To determine when these ANPs reach a steady state before saturation (i.e., bright state), we measured the population of $^4F_{5/2}$ excited energy level by modulating the 1064 nm pump at 5 Hz (Supplementary Figure S21B). Rise times of ca. 20 ms are measured in this case, shorter than for other ANPs or $KPb_2Cl_5:Nd^{3+}$ ANPs at RT[1].

Consequently, PA leading to the IOB is inherently faster than energy-looping processes reported in other ANPs. Optical bistable ANPs directly reach saturation at the threshold value when switched from a dark to a bright state without slow, intermediate photoluminescence.

## 13. Controlled photoswitching of the bistable $KPb_2Cl_5:Nd^{3+}$ nanocrystals

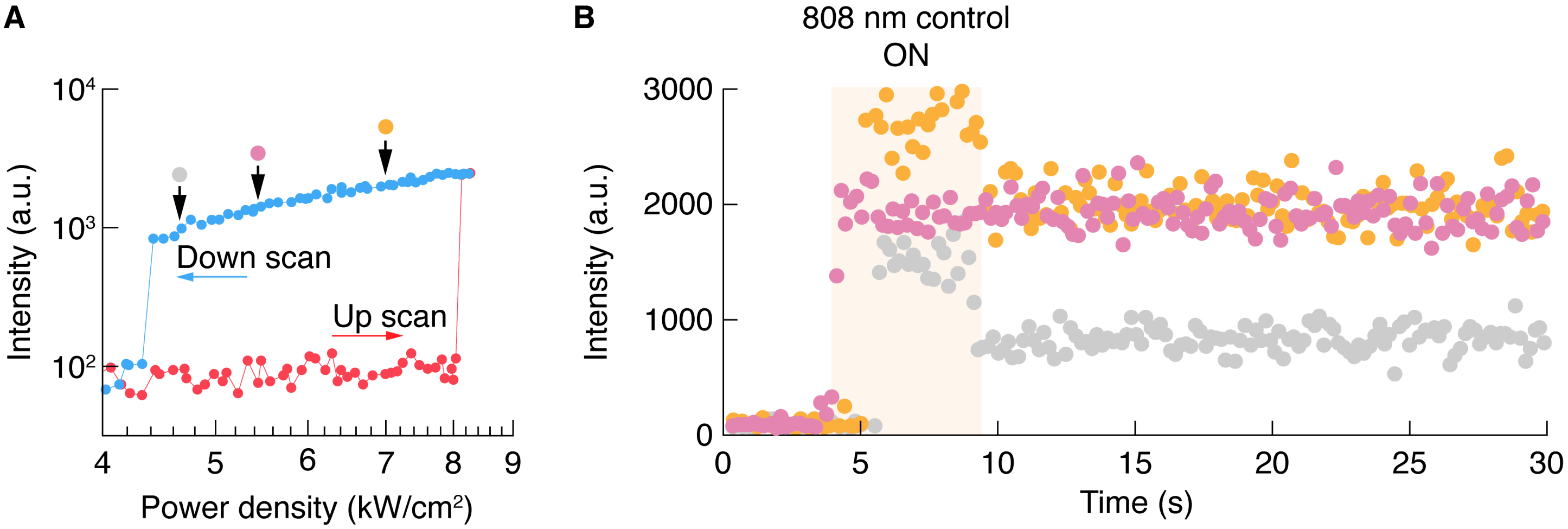

**Supplementary Figure S22. A** – Pump (**1064 nm, 40% duty cycle, 1000 Hz**) power dependence of the bistable $KPb_2Cl_5:Nd^{3+}$ nanocrystals used for photoswitching experiments with the indicated bias at different average power densities. **B** – Emission at 880 nm of the bistable $KPb_2Cl_5:Nd^{3+}$ nanocrystals before and after 808 nm control input. The increase in intensity when the 808 nm control input was switched ON is due to the control laser bleeding into the APD.

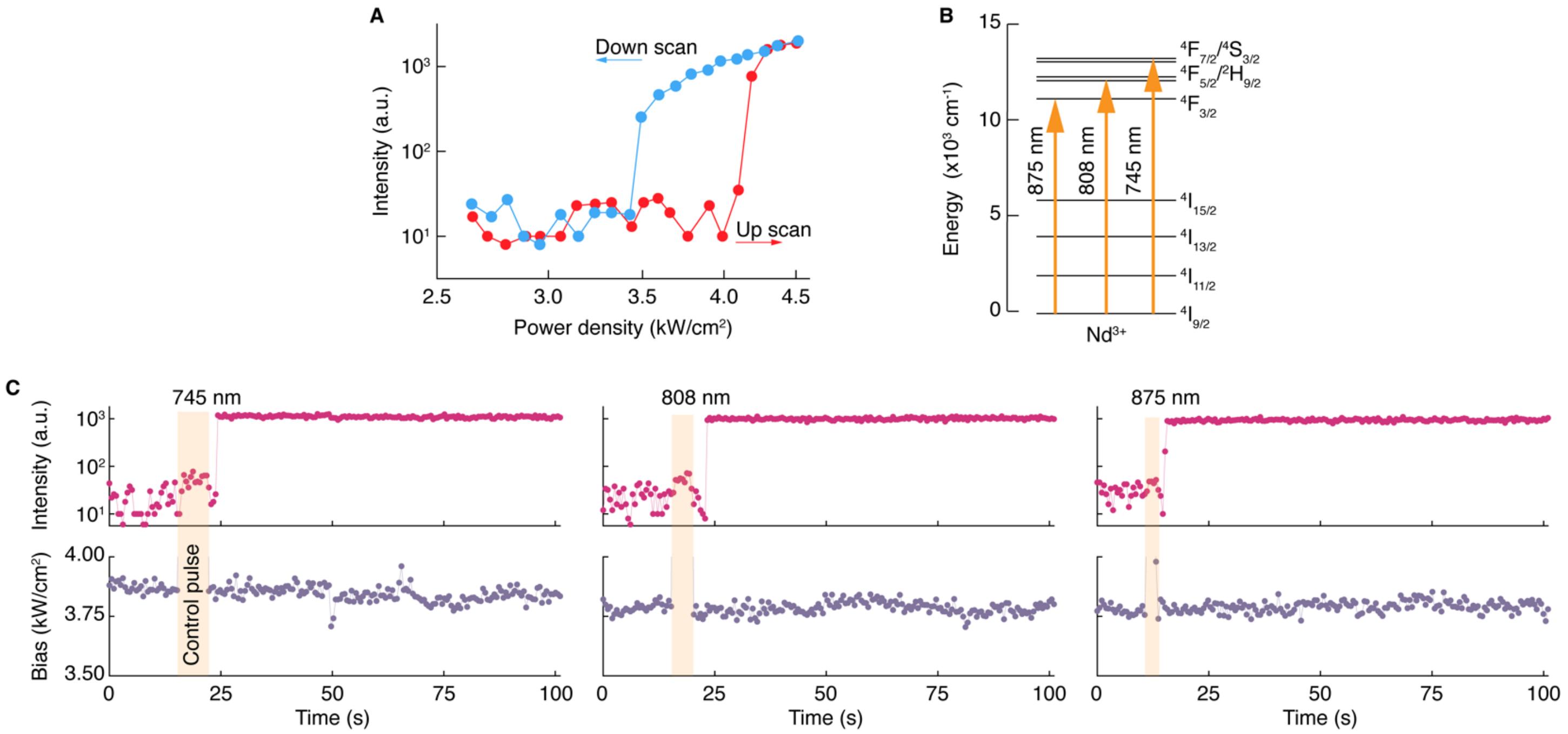

**Supplementary Figure S23**. **A** – Pump (1064 nm) power dependence of the bistable $KPb_2Cl_5:Nd^{3+}$ nanocrystals used for photoswitching experiments. **B** – A simplified $Nd^{3+}$ energy level diagram with the indicated wavelengths of the control laser, matching ground state absorption of $Nd^{3+}$ ions. **C** – time-dependent recording of 1064 nm bias power density and emission at 880 nm of the bistable $KPb_2Cl_5:Nd^{3+}$ nanocrystals before and after exposure to 745, 808, or 875 nm control pulse. Yellow shading indicates when the control input was switched ON and bled into the power meter. The delay in the emission detection is due to the shutter opening in the emission pathway after the control input ceases, which was used to avoid control laser bleeding into the APD.

## 14. Supplementary Discussion S1. Origins of IOB.

Our manuscript uses multiple lines of experimental and theoretical evidence to support our conclusion that IOB in $KPb_2Cl_5:Nd^{3+}$ nanocrystals is, in fact, non-thermal and distinct from the bistability of other $Ln^{3+}$-doped materials. While it is impossible to prove a mechanism, we rule out the most likely alternate hypotheses below.

This evidence includes:

1. Rate equation modeling reproduces the bistable photoluminescence of $KPb_2Cl_5:Nd^{3+}$ without any thermal inputs. To counter the hypothesis that our models include artifacts that lend themselves to IOB, we note that rate equation models developed independently from us have also been used to predict (but not experimentally realize) IOB in a $Yb^{3+}$, $Tm^{3+}$ doped system without incorporating thermal effects[39].
2. If IOB were thermal in nature (e.g., beam-induced heating), one would expect that scanning through different excitation powers at different rates would alter the hysteresis and IOB, since the nanocrystals would experience more heating at longer dwell times. However, we show that this alternate thermal mechanism is unlikely because we still observe IOB of $KPb_2Cl_5:Nd^{3+}$ nanocrystals at different scan rates (Supplementary Figure S6), with little change in hysteresis width and with high reproducibility under continuous and pulsed laser irradiation (Supplementary Figure S16). One would also expect that heating would be dissipated under pulsed excitation at a low duty cycle, yet we still observe IOB in our system even when the laser is pulsed at a 10% duty cycle (Supplementary Figure S17).
3. If IOB were thermal in nature, requiring a full power sweep to trigger delayed avalanching, one would not expect to be able to photoswitch $KPb_2Cl_5:Nd^{3+}$ nanocrystals optically using a second, low-power laser (e.g. 808 nm), without going through a full power sweep. However, the transistor-like performance of $KPb_2Cl_5:Nd^{3+}$ nanocrystals shown in Fig. 5 and Supplementary Figures S22-23 demonstrates that we can indeed access all-optical photoswitching, allowing us to reject the thermal mechanism.
4. If the IOB were thermal in nature, one would expect that it would be highly dependent on the substrate since its thermal properties dictate the dissipation of heat. However, as we demonstrate in the Supplementary Figure S5, the IOB of $KPb_2Cl_5:Nd^{3+}$ nanocrystals does not depend on the substrate on which the nanocrystals were deposited, with luminescence hysteresis obtained on both glass and silicon substrates.
5. A thermal mechanism would require phonon emission to outcompete PA emission. However, multiphonon-relaxation rates are suppressed by the low-phonon-energy $KPb_2Cl_5$ host, while PA is highly efficient, especially at saturation where theoretical quantum yields can approach 40-50% (ref: 40).
6. As a final alternate mechanism, we consider that IOB might occur due to thermal nonlinearities driven by resonant ground state absorption (GSA) of the pump laser[12]. However, our PA nanocrystals are excited at a laser wavelength non-resonant with the GSA, reducing thermal loading and instead producing nonlinearity via electronic cross-relaxation processes. The non-resonant GSA is phonon-assisted, thus the efficient ESA+CR looping in PA effectively bypasses this GSA process. In fact, if thermal processes promoted GSA, this would be expected to decrease the nonlinearity of PA luminescence since PA requires a high ratio (~10,000) of ESA rate to GSA rate. Since we still see extreme nonlinearities (200 and greater), we can dismiss this alternate mechanism.

Having ruled out the most likely alternate hypotheses and with rate equation modeling supporting the most simple explanation — an all-optical mechanism — we believe that the preponderance of the evidence, taken together, supports our assignment of a non-thermal IOB mechanism.